\documentclass[nologo,nourl,11pt,a4paper]{ETHpaper}
\pdfoutput=1
\usepackage{graphicx, epsfig}
\usepackage[numbers]{natbib}
\usepackage{graphicx}                  
\usepackage[x11names, rgb]{xcolor}
\definecolor{col}{rgb}{0.5 0.5 1}
\definecolor{colfail}{rgb}{1 0.5 0.5}

\usepackage{amsfonts}
\usepackage{array}
\usepackage{amsmath}

\begin{document}
\newcommand{\sign}{\mathrm{sign}}
\newcommand{\Prob}{\mathrm{Pr}}
\newcommand{\R}{\mathbb{R}}
\newcommand{\N}{\mathbb{N}}
\newcommand{\nbin}[1]{\textrm{nb}_\textrm{in}(#1)}
\newcommand{\nbout}[1]{\textrm{nb}_\textrm{out}(#1)}
\newcommand{\one}{\mathbb{\bf 1}}
\newcommand{\mean}[1]{\left\langle #1 \right\rangle}
\newcommand{\abs}[1]{\left| #1 \right|}
\newcommand{\comment}[1]{\vspace{0.5cm}\textit{#1}\vspace{0.5cm}}

\title{Systemic Risk in a Unifying Framework for Cascading Processes on
  Networks}

\titlealternative{Systemic Risk in a Unifying Framework for Cascading
  Processes on Networks}

\author{Jan Lorenz, Stefano Battiston, Frank Schweitzer}

\authoralternative{Jan Lorenz, Stefano Battiston, Frank Schweitzer}

\address{ Chair of Systems Design, ETH Zurich, Kreuzplatz 5, 8032 Zurich,
  Switzerland }

\reference{European Physical Journal B, vol. 71, no. 4 (2009), pp. 441--460 \\
  See \texttt{http://www.sg.ethz.ch} for more information.}

\www{\texttt{http://www.sg.ethz.ch}}

\date{\today}

\makeframing
\maketitle

\begin{abstract}
We introduce a general framework for models of cascade and
  contagion processes on networks, to identify their commonalities and
  differences. In particular, models of social and financial cascades, as
  well as the fiber bundle model, the voter model, and models of epidemic
  spreading are recovered as special cases. To unify their description,
  we define the net fragility of a node, which is the difference between
  its fragility and the threshold that determines its failure. Nodes fail
  if their net fragility grows above zero and their failure increases the
  fragility of neighbouring nodes, thus possibly triggering a cascade. In
  this framework, we identify three classes depending on the way the
  fragility of a node is increased by the failure of a neighbour.  At the
  microscopic level, we illustrate with specific examples how the failure
  spreading pattern varies with the node triggering the cascade,
  depending on its position in the network and its degree.  At the
  macroscopic level, systemic risk is measured as the final fraction of
  failed nodes, $X^\ast$, and for each of the three classes we derive a
  recursive equation to compute its value.
  The phase diagram of $X^\ast$ as a function of the initial conditions,
  thus allows for a prediction of the systemic risk as well as a
  comparison of the three different model classes. We could identify
  which model class lead to a first-order phase transition in systemic
  risk, i.e. situations where small changes in the initial conditions may
  lead to a global failure.  Eventually, we generalize our framework to
  encompass stochastic contagion models. This indicates the potential for
  further generalizations.

\end{abstract}

\section{Introduction}
\label{sec:introduction}
After the spread of the financial crisis in 2008, the term 'systemic
risk' could be well regarded as the buzzword of these years. Although
there is no consensus on a formal definition of systemic risk, it usually
denotes the risk that a whole system, consisting of many interacting
agents, fails. These agents, in an economic context, could be firms,
banks, funds, or other institutions.  Only very recently, financial
economics is accepting the idea that the relation between robustness of
individual institutions and systemic risk is not necessarily
straightforward \cite{Morris.Shin2008FinancialRegulationin}.  The debate
on systemic risk, how it originates and how it is affected by the
structure of the networks of financial contracts among institutions
worldwide, is only at the beginning
\cite{Brunnermeier2008Deciphering2007-08Liquidity,lorenz.ea08}. From the
point of view of economic networks, systemic risk can even be conceived
as an undesired externality arising from the strategic interaction of the
agents \cite{schweitzerea09}. However, systemic risk is not only a
financial or economic issue, it also appears in other social and
technical systems. The spread of infectious diseases, the blackout of a
power network, or the rupture of a fiber bundle are just some examples.
Systemic risk -- in our perspective -- is a macroscopic property of a
system which emerges due to the nonlinear interactions of agents on a
microscopic level. As in many other problems in statistical physics, the
question is how such a macroscopic property may emerge from local
interactions, given some specific boundary conditions of the system. The
main research question is then to predict the fraction of failed nodes
$X$ in a system, either as a time dependent quantity or in
equilibrium. Here, we regard $X$ as a measure of systemic risk.

In this paper we investigate systemic risk from a complex network
perspective. Thus, agents are represented by nodes and interactions by
directed and weighted links of a network. Each of the nodes is
characterized by two discrete states $\{0,1\}$, which can be interpreted
as a susceptible and an infected state or, equivalently, as a healthy and
a failed state. In most situations considered here, the failure
(infection) of a node exerts some form of stress on the neighbouring
nodes which can possibly cause the failure (infection) of the neighbours,
this way triggering a cascade, which means that node after node fails.
This may happen via a redistribution mechanism, in which part of the
stress acting on a node is transferred to neighboring nodes, which
assumes that the total stress is conserved. There is another mechanism,
however, where no such conserved quantity exist, for example in infection
processes where the disease can be transferred to an unlimited number of
nodes. In both mechanisms, the likelihood that a node fails increases
with the number of failures in the proximity of the node. This is the
essence of a contagion process.  The specific dynamics may vary across
applications, nevertheless there are common features which should be
pointed out and systematically investigated. Our paper contributes to
this task by developing a general framework which encompass most of the
existing models and allows to classify cascade models in three different
categories.

A number of works have investigated processes of this type, sometimes
referred to as 'cascades' or 'contagion'. These were mostly dealing with
interacting units with random mixing or, more recently, with fixed
interaction structures corresponding to complex networks.  On the one
hand, there are models in which the failure dynamics is deterministic but
the threshold, at which such a failure happens, is heterogeneous across
nodes. For simplicity, we refer to these as \textit{cascade models} --
even though, according to the discussion above, they also involve
contagion. To this class belong some early works on electrical breakdown
in random networks \cite{Kahng.Batrouni.ea1988Electricalbreakdownin} and
more recent ones on the fiber bundle model (FBM)
\cite{sornette1998scaling,moreno2002instability,Kim.Kim.ea2005UniversalityClassof}, on
fractures \cite{Crucitti.Latora.ea2004Modelcascadingfailures}, cascades
in power grids \cite{Carreras.Lynch.ea2004Complexdynamicsof}, or cascades
in sand piles -- the Bak-Tang-Wiesenfeld model (BTW)
\cite{Goh.Lee.ea2003SandpileScale-FreeNetworks}.  Further work refers to
congestion dynamics in networks,
\cite{Bianconi2004Cloggingandself-organized}, cascades in financial
systems \cite{Battiston.Gatti.ea2007CreditChainsand} and in social
interactions \cite{Watts2002SimpleModelof}, and overload distribution (in
abstract terms) \cite{Motter2004CascadeControland}. The properties of
self-organized criticality of some of these models are well understood
\cite{Vespignani1998Howself-organizedcriticality,caruso2006olami}. The
presence of rare but large avalanches is of course relevant to systemic
risk \cite{sornette2009dragon}.

On the other hand, there are models in which the failure of a given node
is stochastic but the threshold at which contagion takes place is
homogeneous across nodes. For simplicity, we refer to this class as
\textit{contagion models}, even though they can lead to cascades as well.
The best known example is epidemic spreading (SIS)
\cite{Pastor-Satorras.Vespignani2001EpidemicSpreadingin}
\cite{Vespignani.Pastor-Satorras2002EpidemicSpreadingScale-free}.  The
properties of these model have been investigated in great detail on
various network topologies, e.g. in the presence of correlations
\cite{Bogun'a2003Absenceofepidemic} or bipatite structure
\cite{gomez2008spreading}. However, as we will see later, we can also
include the voter model (VM) and its variants
\cite{Stark2008Deceleratingmicrodynamicscan,Schweitzer2009Nonlinearvotermodels:}
into this class. It is interesting to note that, while the macroscopic
behaviour of FBM and BTW in a scale free topology is qualitatively
similar to the one on regular and random graphs, the properties of SIS
are severely affected by the topology. The relation between cascading
models and contagion models has not been investigated in depth, although
some models interpolating between the two classes have been proposed
\cite{Dodds2009Analysisofthreshold,Dodds2004Universalbehaviorin}

To relate these two model classes of cascades and contagion, in the
following we develop a general model of cascades on networks where nodes
are characterized by a two continuous variables, \emph{fragility} and
\emph{threshold}. Nodes fail of their fragility exceed their individual
heterogeneous threshold. The key variable is the net fragility $z$,
i.e. the difference between fragility and threshold. This variable is
related to the notion of 'distance to default' used in financial
economics \cite{Avellaneda.Zhu2001Distancetodefault}. By specifying the
the fragility of a node in terms of other nodes fragility and/or other
nodes failure state, we are able to recover various existing cascade
models. In particular, we identify three classes of cascade models,
referred to as `constant load', `load redistribution', `overload
redistribution'. The three classes differ, given that a node fails, in
how the increase in fragility (called here the `load') of connected nodes
is specified. We discuss the differences and similarities among these
classes also with respect to models from financial economics and
sociology. For all of the three classes we derive mean-field recursive
equation for the asymptotic fraction of failed nodes, $X^\ast$. Clearly,
this variable depends on the initial distributions of both fragility and
threshold across nodes. For instance, if no node is fragile enough to fail in the
beginning, then no cascade is triggered. We thus compare how different models
behave depending on the mean and variance of the initial distribution of
$z$ across nodes.

As a further contribution, we extend the general framework to encompass
models of stochastic contagion. In such a framework, the failure of a
given node is a stochastic event depending both on the state of
neighbourhood and on the individual threshold. We derive a general
equation for the expected change of the fraction of failed nodes, from
which one can recover the usual mean-field equations of the SIS model,
but interestingly also of the VM, as special cases.

Our work wishes to contribute to a better understanding of the relations
between cascading models, contagion models and herding models on networks,
from the point of view of systemic risk.

\section{A Framework for Deterministic Models of Cascades}
\label{sec:form-model-cont}
In this section we develop a general framework to describe cascading
processes on a network. This framework will be extended in
Sec. \ref{sec:simple-cont-models} to encompass also stochastic contagion
models. On the microscopic side, we characterize each node $i$ of the
network at time $t$ by a dynamic variable $s_i(t) \in \{0,1\}$
characterizing the failure state. The state is $s_i(t)=1$ if the node has
failed and $s_i(t)=0$ otherwise. Other metaphors apply equally well to
our model, e.g. `infected/healthy', `immune/susceptible', or `broken/in
function'. On the macroscopic side, the system state at time $t$ is
encoded in the $n-$dimensional state vector $s(t)$, with $n$ being the
number of nodes. The macrodynamic variable of interest for systemic risk
is the total \emph{fraction of failed nodes} in the system
\begin{equation}
  \label{eq:xt}
  X(t)=\frac{1}{n} \sum_{i=1}^{n} s_{i}(t).
\end{equation}
If values of $X(t)$ close to one are reached the system is prone to
systemic risk. When trajectories always stay close to zero the system is
free of systemic risk. For simplicity, in the following, we will consider models 
which converge in $X(t)$ to stationary states $X^\ast$. So, the \emph{final fraction of failed
nodes} $X^\ast$ is our proxy for the systemic risk of the system.

In order to describe various existing models in a single framework, we
assume that the failure state $s_i(t)$ of each node is, in turn,
determined by a continuous variable $\phi_{i}(t)$, representing the
\emph{fragility} of the node. A node remains healthy as long as
$\phi_i(t) < \theta_i$, where the constant parameter $\theta_{i}$
represents the \emph{threshold} above which the fragility determines the
failure. Conversely, the node fails if $\phi_i(t) \ge \theta$. In other
words,
\begin{align}
  \label{eq:unify}
  s_i(t+1) = \Theta(z_i(t)) \,\, \text{, with }\,
  z_{i}(t)=\phi_{i}(t)-\theta_{i}
\end{align}
where $\Theta$ is the Heaviside function (here meant to be $\Theta(z)=0$
if $z<0$ and $\Theta(z)=1$ if $z\geq 0$). The variable $z_{i}(t)$ is
called \emph{net fragility}. As it is defined as the difference between
fragility and failing threshold its absolute value has the same meaning
of \textit{distance to default} in finance, for $z\le0$ \cite{Avellaneda.Zhu2001Distancetodefault}.
Notice that in the equation above time runs in discrete steps,
consistently with failure being a discrete event. 

This general framework can be applied to different models by specifying
the functional form of fragility. As we will see, depending on the case
under consideration, $\phi_i(t)$ can be a function of the failure state
vector $s(t)$ and some static parameters, such as the network structure
and the initial distribution of stress on the nodes. It can also be a
function of the vector of fragility $\phi(t-1)$ at previous times. The
latter constitutes a coupled system with the vectors $s(t)$ and $\phi(t)$
as state variables. In any case, fragility depends on the current failure
state and determines the new failures at the next time step. Thus,
cascades are triggered by the fact that failures induce other
failures. Specific models will be described in Sec.
\ref{sec:specific-cascade-models}

The interaction among nodes is specified by the (possibly weighted)
adjacency matrix of the network $A\in \mathbb{R}^{n \times n}$, with
$a_{ij}\geq 0$. For specific models some restrictions to the adjacency
matrix may apply, e.g. one may consider undirected links, no self-links
or some condition on the weights. In this framework the adjacency matrix
of the network influences the dynamics only as a static parameter, i.e.,
we do not consider feedbacks from the state of a node on the link
structure as in \cite{Konig.Battiston.ea2008AlgebraicGraphTheory}. 

If we assume a large number of nodes, it makes sense to look at the
distribution of the net fragility $z(t)$, in terms of its density
function $p_{z(t)}$. Then from Eqn. \ref{eq:xt} and \ref{eq:unify} 
it follows that the fraction of failed nodes at the next time step is given by
\begin{equation}
  \label{eq:Xmacro}
  X(t+1) = \int_0^\infty p_{z(t)}(z)dz = 1-\int_{-\infty}^0 p_{z(t)}(z)dz.
\end{equation}

In the cascading process new failures modify over time the values of fragility of other nodes. 
We can also formulate the dynamics in the space of density functions:
\begin{equation}
  \label{eq:F}
  p_{z(t+1)} = \mathcal{F}(p_{z(t)}).
\end{equation}
If we know both the density function $p_{\mathbf{\phi}(t)}$ of the
fragility at time $t$ and the density function $p_{\theta}$ of the
failing threshold, we can write
\begin{align}
  p_{z(t)}(z) &= p_{\mathbf{\phi}(t)-\theta}(z) = (p_{\mathbf{\phi}(t)}\ast
  p_{-\theta})(z) \nonumber \\
  &= \int p_{\mathbf{\phi}(t)}(y)\ast p_{\theta}(y-z) dy
  \label{eq:P}
\end{align}
with `$\ast$' denoting the convolution. The expression above assumes that
fragility and threshold are stochastically independent across nodes.
Depending on the specific model, the functional operator $\mathcal F$, in
Eqn. (\ref{eq:F}), may also include dependencies on other static
parameters. The general idea is to find a density $p_{z^\ast}$ that is an
attractive fix point of $\mathcal F$, so that the asymptotic fraction
of failed nodes $X^\ast$ is obtained via Eqn. \eqref{eq:Xmacro}.

\section{Specific Cascading Models}
\label{sec:specific-cascade-models}
\begin{figure}[htbp]
  \centering
  \includegraphics[width=0.45\columnwidth]{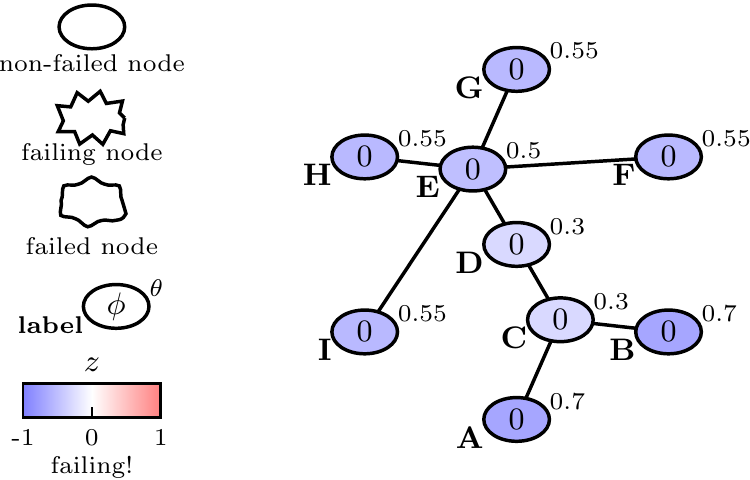}
  \caption{Initial configuration of the generic example used to
    illustrate all models. The legend is valid for all further graphs of
    this type. The discrete state $s_i$ is represented by the shape of
    the node. A healthy node has $s_i=0$, a failed one $s_i=1$. A failing
    node is a node with $s_i=0$ but $z_i>0$, so it will switch to the
    failed state in the next time step. Nodes are labeled with capital
    letters.  The level of fragility $\phi_i$ (which changes over time)
    is indicated inside each node.  The failing threshold $\theta_i$
    (constant over time) is indicated as superscript to the node.  The
    color code specified in the colorbar refers to the value of net
    fragility $z_i=\phi_i-\theta_i$.}
    \label{fig:All:init}
\end{figure}

In many cascading processes on networks, the failure of a node causes a
redistribution of load, stress or damage to the neighbouring nodes. In
our framework, such redistribution of load can be seen as if a failure
causes an increase of fragility in the neighbours. In the following, we
distinguish three different classes of models, denoted as (i) `constant
load', (ii) `load redistribution', and (iii) `overload
redistribution'. We keep the term `load' because it is more intuitive. We
will show how these model classes are described in our unifying framework
in terms of fragility and threshold, and how some models known in the
literature fit into these classes. The differences in the cascading
process across the models will be illustrated by taking the small
undirected network of Figure \ref{fig:All:init} as an example.
For each model, we consider the same initial configuration with respect to
the net fragility $z_i(0)$ in which all nodes are healthy (i.e. with
$z_i(0)$ negative). During the first time step, the value $z_\textbf{C}$ of
node \textbf{C} is perturbed so that it fails. The subsequent time
steps reveal how the propagation of failure occurs in the different
models.

\subsection{Models with Constant Load}
\label{sec:models-with-constant-damage}
Model class (i) (`constant load') assumes that the failure of a node $i$ 
causes a predetermined increase of fragility to
its neighbours.  The term `constant' does not imply that the increase is
uniform for all nodes (on the contrary, some nodes may receive more load
than others). It means that the increase in the fragility of node $i$,
when its neighbor $j$ fails, is the same regardless of the fragility of
$j$ and of the situation in the rest of the system.

We can now distinguish two cases. In the first case, the increase in
fragility of a node $i$ is proportional to the fraction of neighbors that
fail. This is a reasonable assumption if the ties in the network
represent for instance financial dependencies or social influence.
In the second case, the increase in fragility of a node $i$, when
neighbor $j$ fails, is inversely proportional to the number of neighbors
of node $j$. In other words, the load of $j$ is shared equally among the
neighbours and thus the more are its neighbours, the smaller is the
additional load that each one, including $i$, has to carry. We will refer
to the first case as the \textit{inward variant} of the model because the
increase in fragility caused by the failure of one neighbour depends only
on the in-degree of the node receiving the load. In contrast, we will
refer to the second case as the \textit{outward variant}, because the
increase in fragility depends only on the out-degree of the failing node.

We now start by casting in our framework the well known threshold model
of collective behavior by Granovetter
\cite{Granovetter1978ThresholdModelsof}.  The model was developed in the
context of social unrest, with people going on riot when the fraction of
the population which is already on riot exceeds a given individual
activation threshold. This model has been more recently reproposed as
generic model of cascades on networks \cite{Watts2002SimpleModelof}.

We assume an initial vector of failing thresholds $\theta$, and initial
failing states $s_i(0)=0$ for all $i$. We define fragility as simply the
fraction of failed neighbors,
\begin{equation}
  \label{eq:phi:granovetter}
  \phi_i(t) = \frac{1}{k^\textrm{in}_i}\sum_{j\in\nbin{i,A}}s_j(t),
\end{equation}
with $\nbin{i,A}$ being the set of all in-neighbors of $i$ in the network
$A$ and $k^\textrm{in}_i$ being the cardinality of the set (i.e. the  in-degree of
$i$). This means that a node fails when the fraction of its failed neighbors
exceeds its failing threshold. Consequently, the initial fragility across nodes
is zero $\phi_i(0)=0$ for all $i$ and the dynamical equation (\ref{eq:unify}) 
implies $s(1) = \Theta(-\theta)$. Thus, nodes with negative threshold correspond to
initial failures at time step $t=1$.

Interestingly, we can map our inward cascading model with constant load
also to an economic model of bankruptcy cascades introduced in
\cite{Battiston.Gatti.ea2009LiaisonsDangereusesIncreasing}. In that model
firms are connected in a network of credit and supply relations. Each
firm $i$ is characterised by a financial robustness $\rho_i(t)$ which is
a real number, where the condition $\rho_i(t)<0$ determines the default
of the firm. Given a vector $\rho(0)$ of initial values of robustness
across firms and a vector $s(t)$ of failure states, the robustness of
firm $i$ at the next time step is computed as
\begin{equation}
  \rho_i(t+1)=\rho_i^0-\frac{a}{k^\textrm{in}_i}\sum_{j\in\nbin{i,A}}s_i(t)
  \label{eq:robust}
\end{equation}
with $\nbin{i,A}$ being the set of in-neighbors of $i$, $k^\textrm{in}_i$
the in-degree of $i$, and $a$ a parameter measuring the intensity of the
damage caused by the failure. New vectors of failing state vectors and
robustness are then computed iteratively until no new failures occur. Mathematically, 
this process is equivalent to our inward variant model specified 
by Eqn. \eqref{eq:phi:granovetter}. The equivalence is obtained by defining
fragility $\phi_i$ as in Eqn. \eqref{eq:phi:granovetter} and by setting
\begin{equation}
  \theta_i = \frac{\rho^0_i}{a}
  \label{eq:equiv}
\end{equation}
We note that the model specified in
\cite{Battiston.Gatti.ea2009LiaisonsDangereusesIncreasing} also includes
a dynamics on the robustness inbetween two cascades of failures, which
is not part of our framework.

\begin{figure}
  \begin{center}
    \includegraphics[width=0.43\columnwidth]{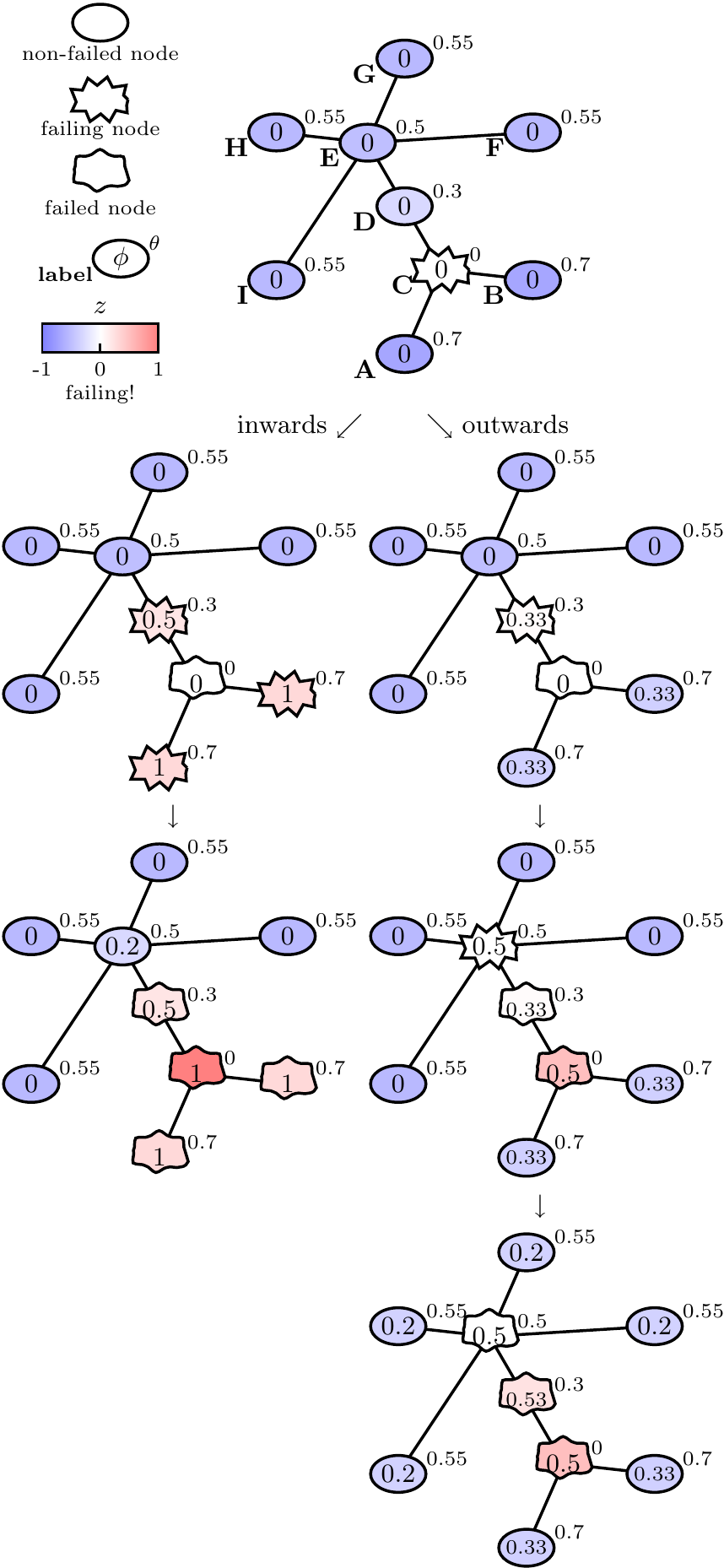}
  \end{center}
\caption{Illustration of the cascading dynamics for the inward (left) and
  outward (right) variants of model class (i) `constant load', based on
  the general example of Figure \ref{fig:All:init}. Initially, node
  \textbf{C} is forced to failure by setting its failure threshold to
  zero. Subsequent time steps in the evolution of the cascade are
  represented downward in the figure.
}\label{fig:FDfail3}
\end{figure}

Let us now turn to the outward variant of the constant load model. It can
be described within our framework by defining fragility as
\begin{equation}
  \label{eq:phi:damagetransfer2}
  \phi_i(t) = \sum_{j\in\nbin{i,A}}\frac{s_j(t)}{k^\textrm{out}_j}
\end{equation}
with $k^\textrm{out}_j$ being the out-degree of node $j$.
If the network is undirected and regular, i.e., all nodes have the same
degree, the inward and the outward model variants
\eqref{eq:phi:granovetter}, \eqref{eq:phi:damagetransfer2} are equivalent
and lead to identical dynamics. However, if the degree is heterogeneous,
then the number and the identity of the nodes involved in the cascade
differ, as shown in the example of Figure \ref{fig:FDfail3}.

Notice that the influence of high and low out-degree nodes interchange in
the two variants, as well as the vulnerability of high and low in-degree
nodes. In the inward variant, high in-degree nodes are more protected
from contagion as they only fail when many neighbours have failed. In
turn, when a high out-degree node fails, it causes a big damage if it has
many neighbors with low in-degree. In contrast, in the outward variant, a
failing low out-degree node generates a larger impact on its neighbours
since the load is distributed among fewer nodes. Thus, a high in-degree
node is more exposed to contagion if it is connected to low out-degree
nodes. On the other hand, a failing high out-degree node does not cause
much damage to its neighbors because the damage gets divided between many
nodes. In the examples reported in the figures, the network is undirected
and in-degree and out-degree coincide. Still the roles of high-degree
and low-degree nodes interchange as discussed above.

As another important difference between the two variants, the maximal
fragility is bounded by the value one in the inward variant, while it is
bounded by the number of nodes $n$ in the outward variant, which is
realized in a star network. Further, both variants strongly differ
regarding the impact of the position of the initial failure. Figure
\ref{fig:FDfail9} (in Appendix \ref{sec:further-examples}) shows an
example, where node \textbf{I} initially fails (instead of node
\textbf{C} in Figure \ref{fig:FDfail3}). The cascade triggered by that
event is larger in the outward variant than in the inward variant, in
contrast to what seen in Figure \ref{fig:FDfail3}. Eventually, Figure
\ref{fig:FDfail5} illustrates the dynamics of a cascade triggered by the
failure of node \textbf{E}, which has the highest degree. This results in a full
cascade in the inward variant, while there is no cascade at all in the
outward variant. This observation illustrates the different influence of
nodes with high degree in the inward and the outward variant, as
explained above.

\subsection{Models with Load
  Redistribution}\label{sec:models-with-redistr-of-load}

Model class (ii) `load redistribution' is our second class of cascading
models. In this class all nodes are initially subject to a certain amount
of load. Actually, in this model class fragility coincides with load. 
When a node $i$ fails, all of its load is redistributed among the first
neighbours. This mechanism differs from class (i) because in class (ii)
the increase in fragility among the neighbours of $i$ depends on the
actual value of $i$'s fragility and not just on the fact that it exceeds
the threshold. The damage caused by one failure can thus not be specified
a priori.

Models belonging to this class include the fiber bundle model (FBM)
\cite{Kun2000Damageinfiber} and models of cascades in power grids
\cite{Kinney2005Modelingcascadingfailures}. In some cases it is possible
to define the total load of the system, which, additionally, but not
necessarily, may be a conserved quantity. For instance, in the FBM a
constant force is applied to a bundle of fibers each of which is
characterized by a breaking threshold. When a fiber breaks, the load it
carries is redistributed equally to all the remaining fibers, so the
total load is conserved by definition. In the context of networks, a node
represents a fiber and if the node fails the load is transferred locally
to the first neighbours in the network. An analogy to power grids is also
possible, with nodes representing power plants, links representing
transmission lines, fragility representing demand and threshold
representing capacity, respectively.

There are, several ways to specify the mechanism of local load
transfer. A first variant is the FBM with local load sharing (LLS) and
load conservation, investigated in
\cite{Kim.Kim.ea2005UniversalityClassof}. We refer to this variant as
LLSC. Despite the fact that load sharing is local, total load is strictly
conserved at any time, due to the condition that links to failed nodes
remain able to transfer load (in other words, links do not fail).  A
second variant implies load shedding instead, and we refer to it as LLSS.
In this variant, all links to failed nodes are removed and the load of a
failing node is transferred only to the first neighbours that are not
about to fail. These are the nodes that are healthy and below the
threshold and thus will be still alive at the next time (although they
may reach the threshold meanwhile). However, if there are no surviving
neighbours, the load is eventually lost (or shed).

In the first variant we can cast the FBM-LLS
\cite{Kim.Kim.ea2005UniversalityClassof} and extend it
to the case of heterogeneous load and directed networks.

From now on, we interpret `load' as `fragility', and `capacity' as
'failing threshold'. Let $\phi^0 \in\R^n$ be the vector of initial
fragility (corresponding to the initial load carried by each node), and
$\theta$ the vector of failing thresholds (or maximal capacity).
(For comparison: In \cite{Kim.Kim.ea2005UniversalityClassof}
the threshold $\theta_i$ for node $i$ is denoted by
$\sigma^\textrm{th}_\nu$ with values taken from a uniform distribution
between zero and one.  The load of each node is the same and called
$\bar\sigma = \tfrac{\sigma}{n}$, with $\sigma$ being the total load.)

We define
\begin{equation}
 \mathrm{reach}^{1\to 0}_\mathrm{out}(i,s,A) = \{j \,|\, s_j=0 \text{, $\exists$
path of $1$-nodes $i\to j$} \}
\end{equation}
as the set of healthy nodes which are reachable from node $i$ following
directed paths consisting only of failed nodes (except $i$). Let
$k^{\textrm{reach}_\textrm{out}}_i$ be the cardinality of such
set. Moreover, we define
\begin{equation}
 \mathrm{reach}^{0\to 1}_\mathrm{in}(i,A,s) = \{j \,|\, s_j=1 \text{, $\exists$
path of $1$-nodes $i\to j$} \}
\end{equation}
to be the set of nodes from which node $i$ can be reached along directed
paths consisting of failed nodes (except $i$).  Both sets of nodes
defined above have to be computed dynamically based on the current vector
of failing states $s$ and the network.

Finally, given the initial fragility vector $\phi^0$, the failure state
vector $s(t)$, and the network $A$, we define the fragility of node $i$
at time $t$ in the LLSC variant as
\begin{equation}
 \phi_i(t) = \phi^0_i +
\sum\limits_{j\in \mathrm{reach}^{0\to 1}_\mathrm{in}(i,A,s)}
\frac{\phi_j^0}{\#\mathrm{reach}^{1\to
0}_\mathrm{out}(j,s,A)}.\label{eq:phi:fiberLLSC}
\end{equation}
We add that for an undirected network and uniform initial load, such a
definition becomes equivalent to the \emph{load concentration} factor of
node $i$, as defined in \cite{Kim.Kim.ea2005UniversalityClassof}.

The assumption that links do not break and remain able to transfer load
is not always satisfactory. Some models have thus investigated the LLSS
variant of the model in which the load is transferred only to the
surviving first neighbours \cite{moreno2002instability}. In this case the
load transfer is truly local and there is no transmission along a chain
of failed nodes. This implies that during a cascade of failures, at some
point in time the network might split into disconnected components which
cannot transfer load to each other. In particular, if one of these
subnetworks fails entirely, all the load carried by this subnetwork is
shed.

As a consequence of the LLSS assumption, fragility now is not just a
function of the current state vector $s(t)$ and some static parameters
(such as the network matrix and the initial fragility $\phi^0$). In
contrast, it has to be defined through a dynamic process as a function of
the fragility vector at previous time $t$, according to the following
equation:
\begin{align}
\label{eq:phi:fiberLLSS}
 \phi_i(t+1) &= \left\lbrace
   \begin{array}{cl}
      \phi_i(t) + \sum\limits_{j \in \mathrm{fail}_\mathrm{in}(i)}
\frac{\phi_j(t)}{\#\mathrm{hea}_\mathrm{out}(j)} & \ \text{if
  $\begin{array}{l} s_i(t)=0, \\ \phi_i(t)<\theta_i\end{array}$}
\\
      0 & \ \text{otherwise,}
    \end{array}
  \right. 
\end{align}
with $\mathrm{fail}_\mathrm{in}(i)$ being the set of in-neighbors of $i$
which fail at time $t$ (but have not already failed!), and
$\mathrm{hea}_\mathrm{out}(j)$ the set of out-neighbors of $j$ which
remain healthy at time $t+1$
\begin{align}
\mathrm{fail}_\mathrm{in}(i) &= \{j \,|\, j\in\mathrm{nb}_\mathrm{in}(i,A),
s_j(t) = 0, \phi_j(t)\geq\theta_j\}, \nonumber\\
\mathrm{hea}_\mathrm{out}(j) &= \{i \,|\, i\in\mathrm{nb}_\mathrm{out}(j,A),
s_i(t) = 0, \phi_j(t)<\theta_j\}.
\end{align} 

Thus, Eqn. \eqref{eq:phi:fiberLLSS} is well defined unless
$\mathrm{hea}_\mathrm{out}(j)$ is empty. In this case, there is no
healthy neighbour of $j$ to which the load can be transferred, thus the
load has to be shed. The remaining healthy nodes remain unaffected.

\begin{figure*}
  \begin{center}
    \includegraphics[width=0.9\textwidth]{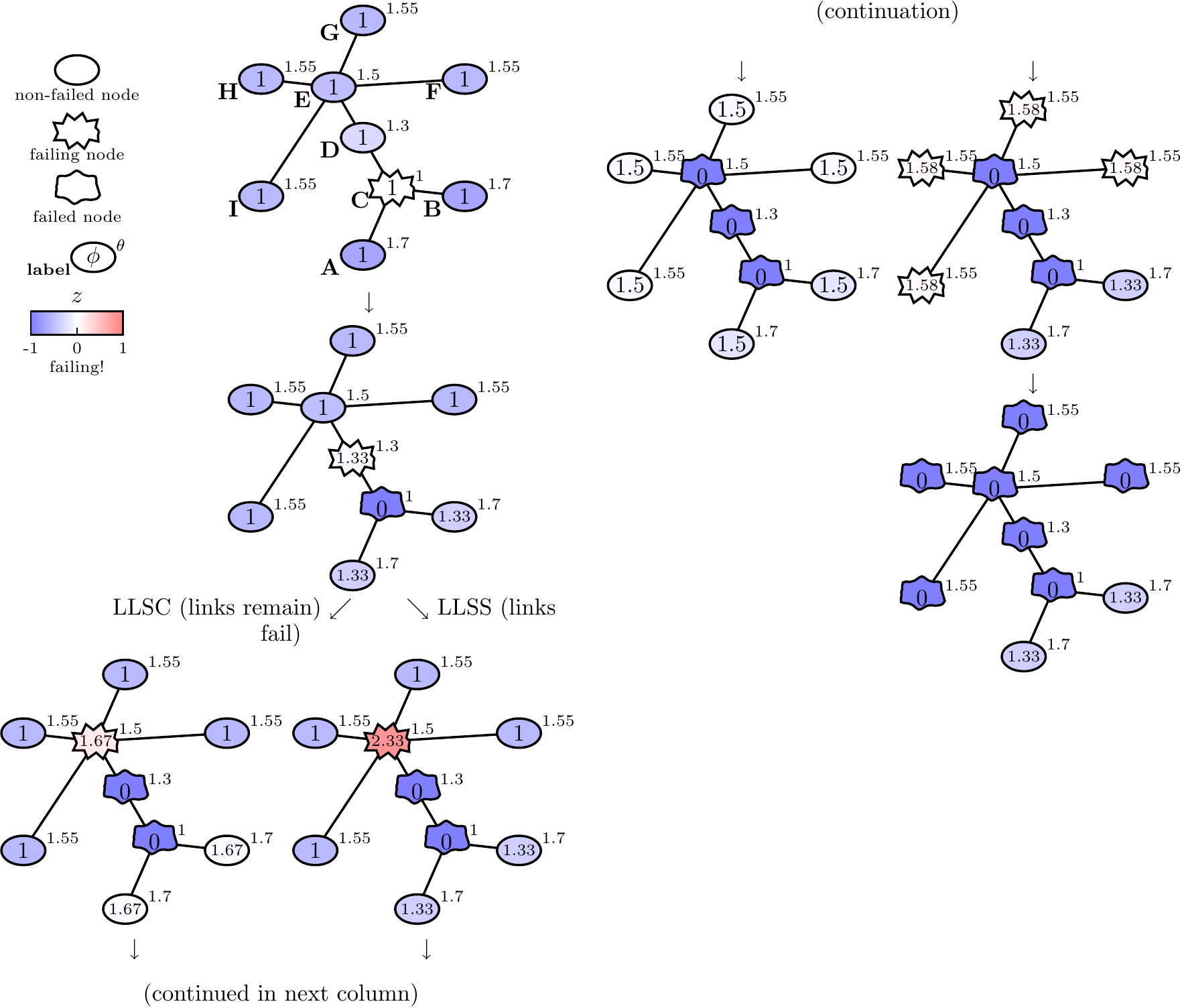}
  \end{center}
\caption{
  Illustration of the cascading dynamics for the two variants of the
  model class (ii) `load redistribution', based on the general example in
  Figure \ref{fig:All:init}. Left: LLSC variant, following Eqn.
  \eqref{eq:phi:fiberLLSC}. Right: LLSS variant following
  Eqn. \eqref{eq:phi:fiberLLSS}. Again, initially node \textbf{C} is
  forced to failure. The dynamics is the same for the two variants in the
  first time step but it differs in the subsequent time steps.
}\label{fig:FBfail3}
\end{figure*}

Figure \ref{fig:FBfail3} illustrates, as an example, the different
outcomes of the dynamics in the LLSC and LLSS variants. The initial load
is set to one for all nodes, thus the total load on the system is
nine. The values of the threshold are set in order to have the same
values of $z=\phi-\theta$ at each node as in the example of Figure
\ref{fig:All:init}. As in Figure \ref{fig:FDfail3}, we set the failing
threshold of node \textbf{C} to one in order to trigger an initial
failure.

On one hand, we could expect that cascades triggered by the failure of
one node are systematically wider in the LLSC variant than in the LLSS
variant because in the first one the total load is conserved. On the
other hand, in the LLSC, the fragility is redistributed also to indirect
neighbours thus leading to a smaller increase of fragility per node and
therefore possibly to smaller cascades. In fact there seems to be no
apparent systematic result, the outcome being dependent on the network
structure and the position of the initial failure. In the example shown in
Figure \ref{fig:FBfail3} the cascade stops sooner in the LLSC variant
than in the LLSS one, due to the rebalancing of load across the
network. In other cases, however, if for instance node \textbf{E} initially
fails, we find that the load shedding has a stronger impact and the
cascade is smaller in the LLSS case.

\subsection{Models with Overload Redistribution}\label{sec:models-with-redistr-overload}
We conclude our classification with class (iii) `overload
redistribution'. When a node $i$ fails in these models, only the
difference between the load and the capacity is redistributed among the
first neighbours. Actually, the overload of a node is its net
fragility. This class is more realistic in applications, where a failed
node can still hold its maximum load and only has to redistribute its
overload.

The Eisenberg-Noe model is an important example of an economic model in
which firms are connected via a network of liabilities
\cite{Eisenberg.Noe2001SystemicRiskin}. When the total liabilities of a
firm $i$ exceed its expected total cash flow (consisting of the operating
cash flow from external sources and the liabilities of the other firms
towards $i$), the firm goes bankrupt. When a new bankruptcy is recognized
the expected payments from others decline, but they do not vanish
entirely. Thus the loss spreading to the creditors is mitigated.

With respect to our framework, we can identify the total liability minus
the currently expected payments (from the liabilities of others) with
fragility. Similarly, operating cash flow corresponds to the failing
threshold. The relation between the Eisenberg-Noe model and the overload
redistribution class is discussed more in detail in Appendix
\ref{sec:eisenbergnoe}.

Therefore, we adapt the two variants of load redistribution defined as LLSC and
LLSS in Section \ref{sec:models-with-redistr-of-load} to the case of
overload redistribution by subtracting the threshold value in the
nominator of Eqns.~\ref{eq:phi:fiberLLSC} and \ref{eq:phi:fiberLLSS}. We
have
\begin{equation}
 \phi_i(t) = \phi^0_i +
\sum\limits_{j\in \mathrm{reach}^{0\to 1}_\mathrm{in}(i,A,s)}
\frac{\phi_j^0-\theta_j}{\#\mathrm{reach}^{1\to
0}_\mathrm{out}(j,s,A)}.\label{eq:phi:overloadLLSC}
\end{equation}
as definition of fragility in the LLSC version when links remain, and
\begin{align}
\label{eq:phi:overloadLLSS}
 \phi_i(t+1) &= \left\lbrace
    \begin{array}{cl}
      \phi_i(t) + \sum\limits_{j \in \mathrm{fail}_\mathrm{in}(i)}
\frac{\phi_j(t)-\theta_j}{\#\mathrm{hea}_\mathrm{out}(j)} & \ \text{if
  $\begin{array}{l} s_i(t)=0, \\ \phi_i(t)<\theta_i\end{array}$}
\\ 
      0 & \ \text{otherwise,}
    \end{array}
  \right. 
\end{align}
as dynamical equation of fragility in the LLSS version when links break. 

\begin{figure}
  \begin{center}
    \includegraphics[width=0.43\columnwidth]{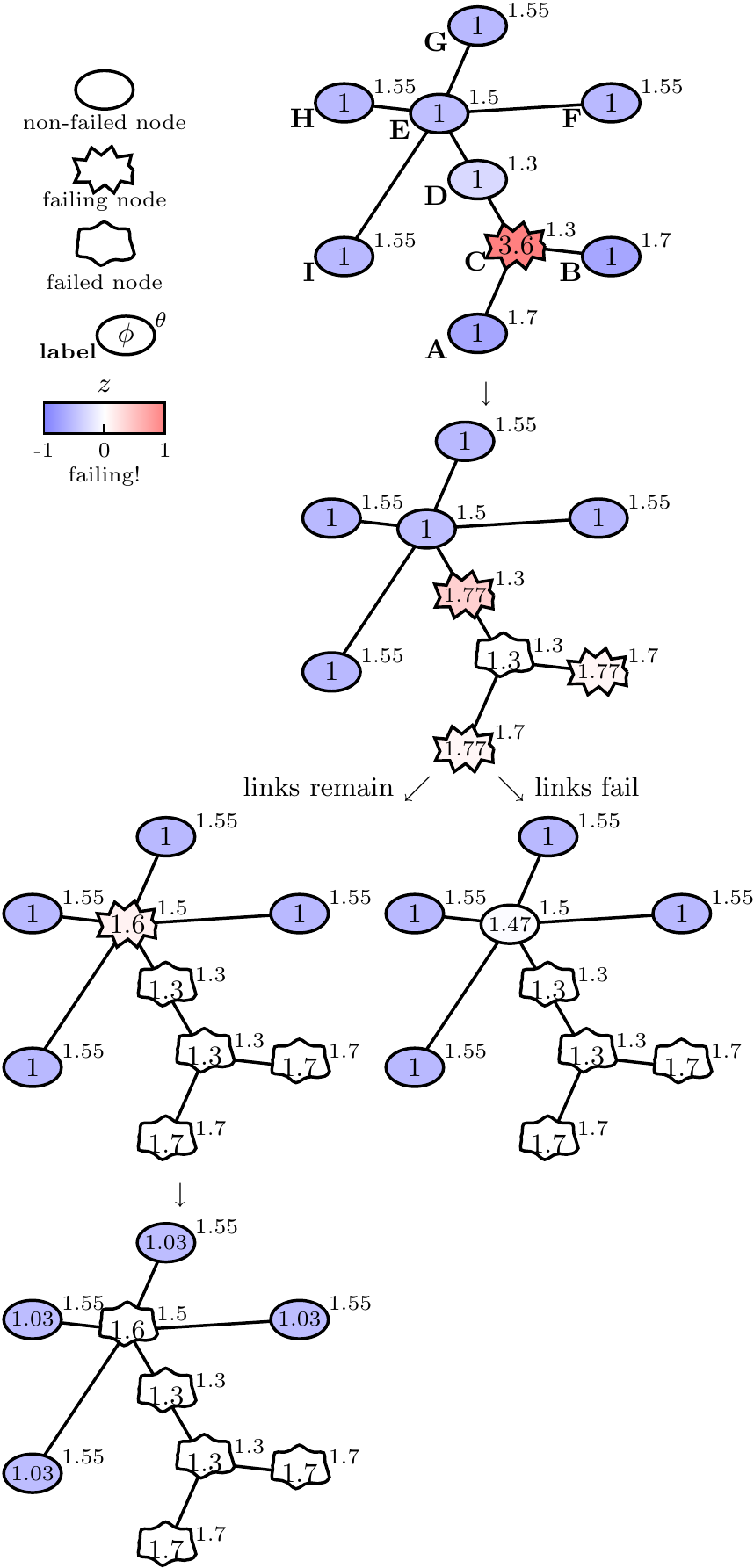}
  \end{center}
 \caption{Illustration of the cascading dynamics for model class (iii)
   `overload redistribution'. Left: LLSC variant based on
   Eqn. \eqref{eq:phi:overloadLLSC}. Right: LLSS variant based on
   Eqn. \eqref{eq:phi:overloadLLSS}.}
\label{fig:FBOfail3}
\end{figure}

Using our small example of Figure \ref{fig:All:init}, the cascading
dynamics for the model class (iii) is presented in Figure
\ref{fig:FBOfail3}. In general, as we will see in Section
\ref{sec:macr-reform} this class of models leads to much smaller
cascades, compared to class (ii). In this example, we have set the
initial fragility of node \textbf{C} high enough so that a large cascade
is triggered.
A very high initial overload is needed to trigger a cascade of failures
because this overload is the only amount which is transferred through the
whole system. On a failure nothing new is added to the total amount,
because the node stays with its maximum capacity.

Notice that the models of overload redistribution are invariant to joint
shifts in the initial fragility $\phi^0$ and in the failing threshold
$\theta$. In other words, a system with $\phi^0+c$ and $\theta+c$ leads
to the same trajectory of failure state $s(t)$ and fragility
$\phi(t)+c$. Thus, it is enough to study the model with $\phi^0=0$
without loss of generality.

\section{Macroscopic reformulations}\label{sec:macr-reform}

In the previous section we have seen that the different classes of
cascading models lead to a diverse behaviour, at least in small scale
examples, even if initial conditions for net fragility are the same. In
this section, by studying simple mean-field approximations of the
processes we find that there are significant differences also at the
macroscopic level. In order to compare the different model classes under
the same conditions, we have set the probability density functions
$p_{z(0)}$ of initial values of the net fragility to be equal for all
models. For the cases (i) constant load, and (iii) overload
redistribution we set $\theta=-z(0)$. Notice, that we can set $\phi^0=0$
in case (iii) without loss of generality. For case (ii) load
redistribution, instead, it is necessary to have $\phi^0 >0$ (otherwise
there is no load to redistribute) and we have $\theta = \phi^0 -
z(0)$. We further assume that the initial fragility $\phi^0$ is uniform
across nodes in model class (ii).

Even a basic mean-field approach allows for an interesting comparison of
the three model classes. To do so, we replace the distribution of
fragility at time $t$, with the delta function $\delta_{\mean{\phi(t)}}$
centered on the mean fragility $\mean{\phi(t)}$. This is equivalent to
assuming a fully connected network since in such a case
Eqns. (\ref{eq:phi:granovetter}-\ref{eq:phi:fiberLLSS}) yield the same
fragility for every node. If the two distributions are independent, from
\eqref{eq:P} we get
\begin{equation}
 p_{z(t)} = \delta_{\mean{\phi(t)}}\ast p_{-\theta}.
\end{equation}
Convolution with a delta corresponds to a shift in the variable, so that
$p_{z(t)} = p_{\mean{\phi(t)}-\theta}$, and from Eqn. \eqref{eq:Xmacro} we
obtain
\begin{align}\nonumber
  X(t+1) &= \int_0^\infty p_{\mean{\phi(t)}-\theta}(z)dz =
  \int_{-\mean{\phi(t)}}^\infty p_{-\theta}(z)dz \\ 
  &= \int_{-\infty}^{\mean{\phi(t)}} p_{\theta}(z)dz =
  P_{\theta}(\mean{\phi(t)})
\label{eq:Xvsmeanphi}
\end{align}
where $P_\theta(x) = \int_{-\infty}^{x} p_{\theta}(\theta)d\theta$ is the
cumulative distribution function of $\theta$.  This is equivalent to a
change of variable $z(t) = \mean{\phi}-\theta$ in the probability
distribution and in the integral. However, the procedure with convolution
can be carried out also if $p_{\phi}$ is not assumed to be a delta
function.

At this point, we have to express the mean fragility $\mean{\phi(t)}$ in
terms of the current fraction of failed nodes, $X(t)$. For case (i)
'constant load', in a fully connected network,
Eqns. (\ref{eq:phi:granovetter}) and (\ref{eq:phi:damagetransfer2}) yield
both the following mean fragility:
\begin{equation}
\mean{\phi(t)} = X(t)
\label{eq:i}
\end{equation}
For case (ii) 'load redistribution', assuming that the surviving nodes
equally share the initial load, we can write for the mean fragility:
\begin{equation}
\mean{\phi(t)} = \frac{\phi^0}{1-X(t)}
\label{eq:ii}
\end{equation}
This is obtained from Eqn. \ref{eq:phi:fiberLLSC} at microscopic level by
taking the mean over all $i$ on both sides
\begin{equation}
 \mean{\phi_i(t)} = \mean{\phi^0_i} + \mean{
\sum\limits_{j\in \mathrm{reach}^{0\to 1}_\mathrm{in}(i,A,s)}
\frac{\phi_j^0}{\#\mathrm{reach}^{1\to
0}_\mathrm{out}(j,s,A)}}
\end{equation}
Now, assuming $\phi^0_i=\phi^0$ for all nodes and the network as fully
connected, we have that: $\mean{\phi^0_i}$ coincides with $\phi^0$; the
sum over the set $\mathrm{reach}^{0\to 1}_\mathrm{in}(i,A,s)$ (which now
coincides with the set of failed nodes) equals $nX(t)$; and
$\#\mathrm{reach}^{1\to 0}_\mathrm{out}(j,s,A)$ equals $n(1-X(t))$,
because we count all healthy nodes. Thus, we obtain
\begin{equation}
  \mean{\phi(t)} = \phi^0 + \frac{nX(t)\phi^0}{n(1-X(t))} = \frac{\phi^0}{1-X(t)}.
\end{equation}
For case (iii) 'overload redistribution', we can proceed similarly
starting from Eqn. \ref{eq:phi:overloadLLSC}. Setting $\phi_i^0=0$,
without loss of generality, and taking the mean over $i$ on both sides
yields
\begin{equation}\label{eq:deriveoverloadmacro}
  \mean{\phi_i(t)} = \mean{
    \sum\limits_{j\in \mathrm{reach}^{0\to 1}_\mathrm{in}(i,A,s)}
    \frac{-\theta_j}{\#\mathrm{reach}^{1\to
        0}_\mathrm{out}(j,s,A)}}
\end{equation}
Again, we can replace the sum over $\mathrm{reach}^{0\to
  1}_\mathrm{in}(i,A,s)$ by $nX(t)$, and $\#\mathrm{reach}^{1\to
  0}_\mathrm{out}(j,s,A)$ by $n(1-X(t))$. However, now the average of the
threshold values $\theta_j$ across all failed nodes (as indicated by the
sum) is not simply $\mean{\theta}$. It is instead the mean of that part
of the distribution $p_\theta$ where failed nodes are located. These are
the nodes with $\theta\le \phi(t)$ and their probability mass has to sum
up to $X(t)$. For a given distribution $p_{\theta}$ and a given fraction
$X$ of failed nodes, the mean threshold of failed nodes is defined as
\begin{equation}
\label{eq:meantheta}
 \mean{\theta}_{X}= \left( \int_{-\infty}^{q_{X}} \theta
 p_{\theta}(\theta) d\theta \right) / X.
\end{equation}
$q_X$ denotes the $X$-quantile of the distribution $p_\theta$, i.e. a
fraction $X$ of the probability mass lies below $q_X$:
\begin{equation}
\label{eq:quantile}
X=\int_{-\infty}^{q_{X}} p_{\theta}(\theta) d\theta
\end{equation}
Thus, $\mean{\theta}_{X}$ is the first moment of $\theta$ below the value
$q_X$, normalized by the probability mass of the distribution
$p_{\theta}$ in the same interval. Replacing this into
Eqn. (\ref{eq:deriveoverloadmacro}) yields as mean fragility for case (iii) overload redistribution:
\begin{equation}
  \quad  \mean{\phi(t)} = \frac{-\mean{\theta}_{X(t)}X(t)}{1-X(t)}.
\end{equation}
Notice, that the mean of the threshold of the failed nodes is negative,
thus the minus in front of $\mean{\theta}_{X(t)}$ ensures that fragility
is positive.

By replacing the expressions of $\mean{\phi(t)}$ in terms of
$X(t)$ in Eqn. (\ref{eq:Xvsmeanphi}) we obtain simple recursive
equations in $X(t)$ for the different cases:
For case (i) 'constant damage'
\begin{equation}
 X(t+1) = P_\theta(X(t)),
\label{eq:recur:FD}
\end{equation}
for case (ii) 'load redistribution'
\begin{equation}
  X(t+1) = P_\theta\left(\frac{\phi^0}{1-X(t)}\right), 
\label{eq:recur:LR}
\end{equation}
and for case (iii) 'overload redistribution'
\begin{equation}
  X(t+1) = P_\theta\left(\frac{-\mean{\theta}_{X(t)}X(t)}{1-X(t)}\right).
\label{eq:recur:OR} 
\end{equation}
As the functions on the right hand sides of Eqns.
(\ref{eq:recur:FD})-(\ref{eq:recur:OR})  are monotonic non-decreasing and
bounded within $[0,1]$, $X(t)$ always converges to a fix point $X^\ast$
representing the final fraction of failed nodes.

\begin{figure}
\centering 
\includegraphics{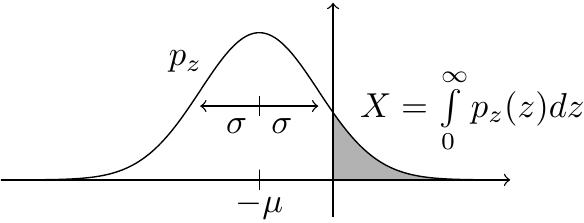} \\
\includegraphics[width=0.29\columnwidth]{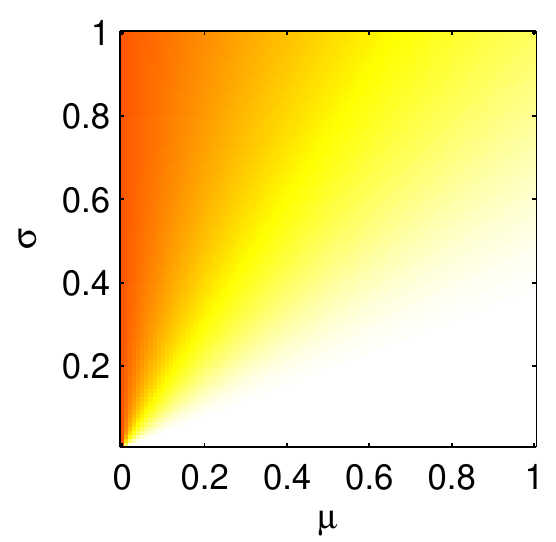} \qquad
\includegraphics[height=0.29\columnwidth]{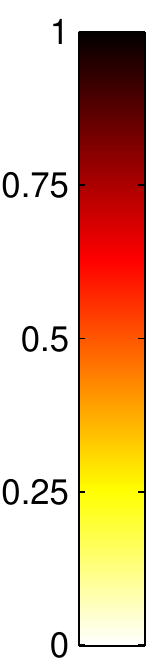} 
\caption{Top: Illustration of the distribution of net fragility and
  geometric interpretation of $X$. Bottom: Initial fraction of failing
  nodes $X(0)$ as a function of mean $-\mu$ and standard deviation
  $\sigma$ of the distribution of initial net fragility
  $z(0)=\phi(0)-\theta$. The distribution is assumed to be normal.}
\label{fig:failfracinit}
\end{figure}

With these iterations we can study the three different models
systematically on the same initial conditions. We assume the failing
thresholds to be normally distributed such that $z(0) \sim
\mathcal{N}(-\mu,\sigma)$ in all three cases. This is guaranteed if
$\theta \sim \mathcal{N}(\mu,\sigma)$ for the cases of constant load and
overload redistribution, and if $\theta \sim
\mathcal{N}(\mu+\phi^0,\sigma)$ for the case of load redistribution. The
parameters $\mu$ and $\sigma$ represent the mean and the standard
deviation of the net fragility. Particularly, $\sigma$ represents the
initial heterogeneity across agents.  The initial fraction of failed
nodes is thus $X(0) = \Phi_{\mu,\sigma}(0)$ where $\Phi_{\mu,\sigma}$
denotes the cumulative distribution function of a normal distribution
with mean $\mu$ and standard deviation $\sigma$. The surface of values
taken by the initial fraction of failed nodes $X(0)$ over the plane
$(\mu,\sigma)$ is shown in Figure \ref{fig:failfracinit}. This is assumed
to be the same in all three cases.

\begin{figure}
\centering 
\includegraphics[width=0.29\columnwidth]{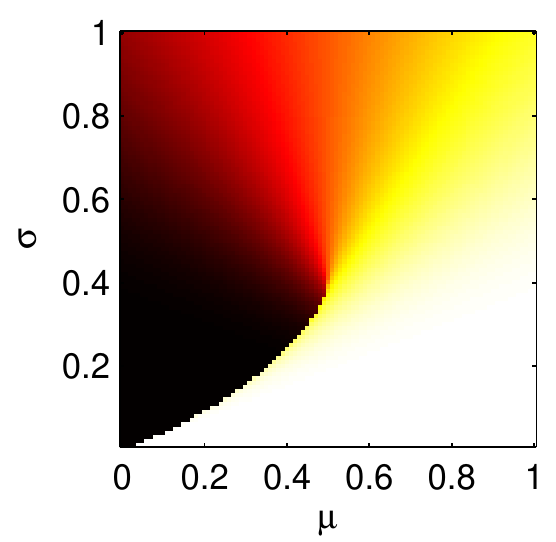}
\includegraphics[width=0.29\columnwidth]{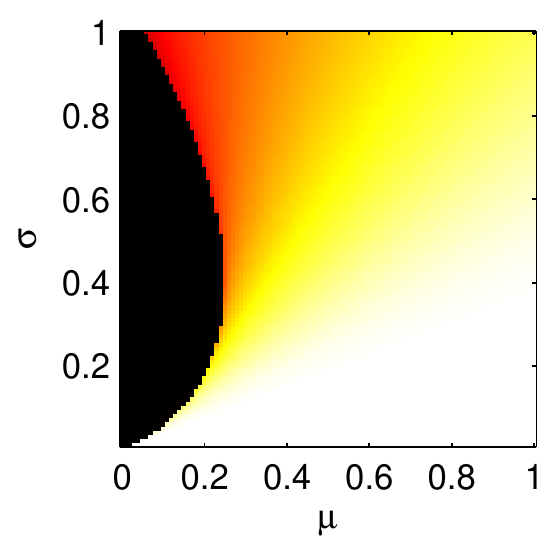}\\
\includegraphics[width=0.29\columnwidth]{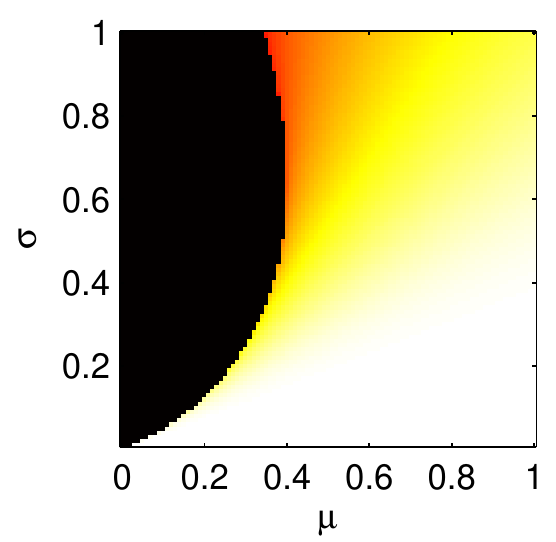}
\includegraphics[width=0.29\columnwidth]{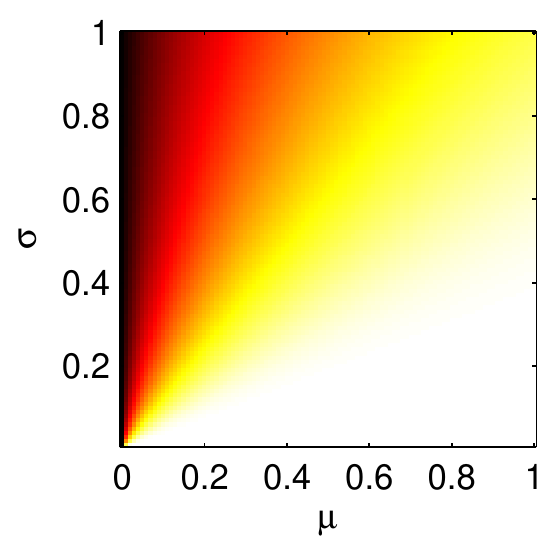}\\
\includegraphics[width=0.29\columnwidth]{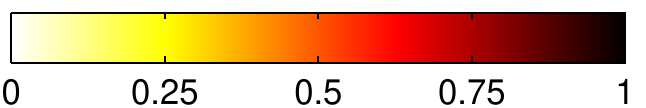} 
\caption{Final fraction of failed nodes $X^\ast$ in mean field
  approximation. As shown in Fig. \ref{fig:failfracinit}, the values
  $-\mu$ and $\sigma$ refer to the initial distribution of
  $z(0)=\phi(0)-\theta$. The various plots refer to the different model
  classes. Top left: class (i) constant load. Top right: class
  (ii) load redistribution with initial load $\phi^0=0.25$. Bottom left:
  class (ii) with $\phi^0=0.4$. Bottom right: class (iii) overload
  redistribution.}
\label{fig:failfrac}
\end{figure}

\begin{figure}
\centering 
\includegraphics[width=0.29\columnwidth]{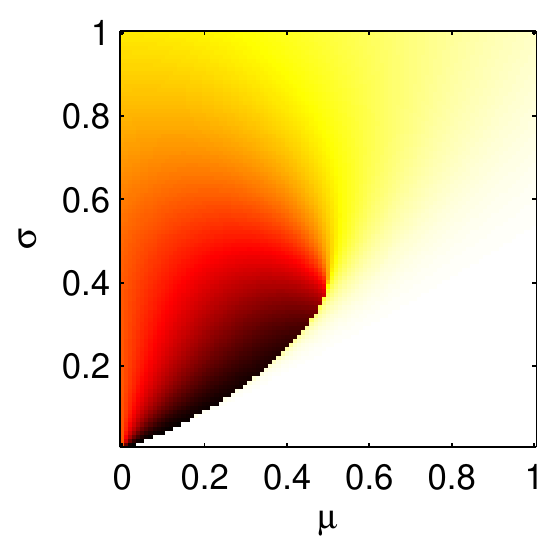}
\includegraphics[width=0.29\columnwidth]{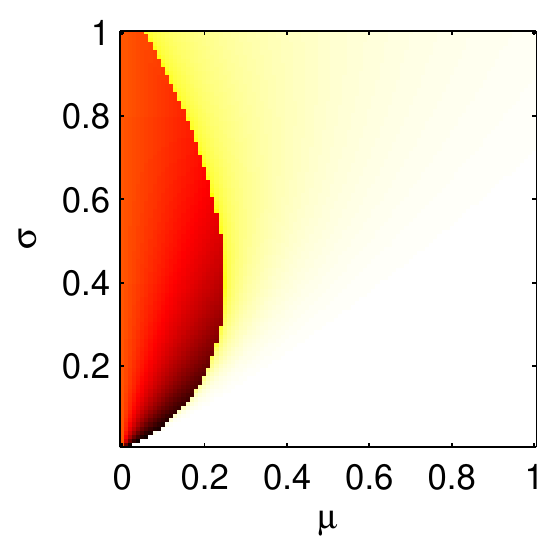}\\
\includegraphics[width=0.29\columnwidth]{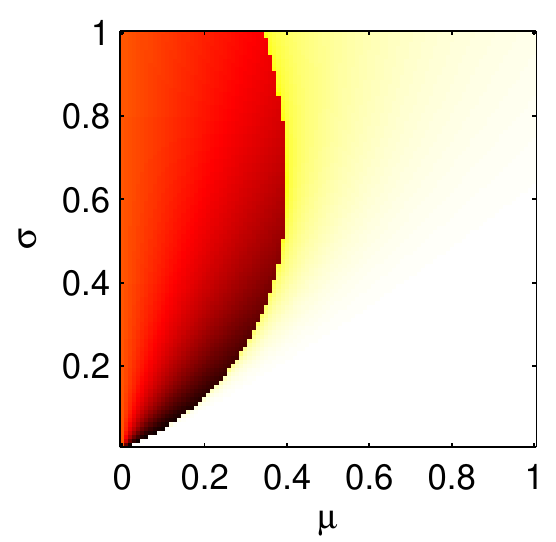}
\includegraphics[width=0.29\columnwidth]{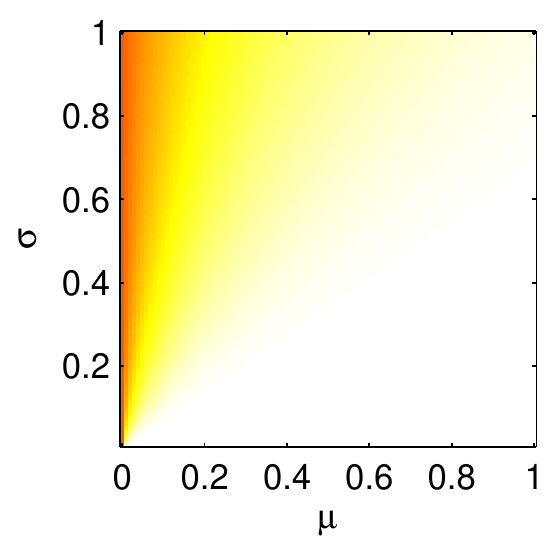}\\
\includegraphics[width=0.29\columnwidth]{nfigs/nfigMacroLegendHorizontal} 
\caption{Net fraction of failed nodes $X^\ast-X(0)$ due to the
  cascading process. Plots are obtained from those in Figure \ref{fig:failfrac}
  by subtracting the initial number of failed nodes (shown in Figure
  \ref{fig:failfracinit}).}
\label{fig:contfailfrac}
\end{figure}

In contrast, the final fraction of failed nodes, $X^\ast$, obtained as
numerical solution of the recursive equations
(\ref{eq:recur:FD})-(\ref{eq:recur:OR}) is shown in Figure
\ref{fig:failfrac}. Moreover, the difference between the two previous
quantities, $X^\ast-X(0)$, representing the fraction of nodes which fail
due to the cascade process, is shown in Figure \ref{fig:contfailfrac}.

At the macro level, the most important structural difference between the
three model classes concerns the existence of a discontinuity and its
boundaries in the landscape of $X^\ast$. Since $X^\ast$ can be considered
an order parameter for our system, regions with different values
separated by a discontinuity indicate a first-order phase transition. The
proximity of the discontinuity, i.e. across the boundary of the phase
transition, marks a region of great interest from the point of view of
systemic risk. Indeed a small change in the distribution of initial net
fragility can mean the diﬀerence between a negligible cascade or a full
breakdown. In the case of 'constant load', we find a discontinuity
between a region with low systemic risk and one with high systemic risk,
with a separation line that starts at $(\mu,\sigma)=(0,0)$ and
monotonically increases in $\mu$ and $\sigma$.  However, above
$(\mu,\sigma)\approx (0.5,0.4)$ the discontinuity vanishes. In the case
of load redistribution, instead, the a region of full break down is
always separated by a discontinuity from the region of partial
survival. Interestingly, the separation line when $\mu$ is seen as a
function of $\sigma$ is not monotonic. For some $\mu$ (e.g $\mu=0.2$ when
$\phi^0=0.25$) most nodes survive for small $\sigma$, half of the nodes
survive for large $\sigma$ but all nodes fail for intermediate
$\sigma$. This means that the system is more robust for low heterogeneity and
high heterogeneity, but more susceptible to systemic risk in the region
of intermediate heterogeneity.
Finally, in the case of overload redistribution region of full breakdown
is reduced to the line with $\mu=0$ with no discontinuity towards the region
of partial survival. 

\begin{figure}
\centering 
\includegraphics[width=0.29\columnwidth]{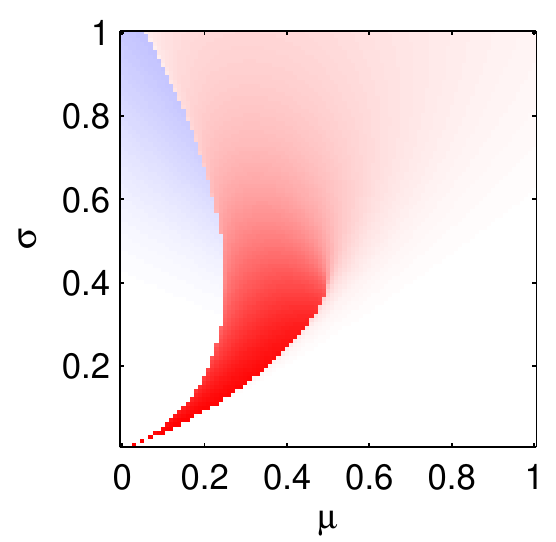}
\includegraphics[width=0.29\columnwidth]{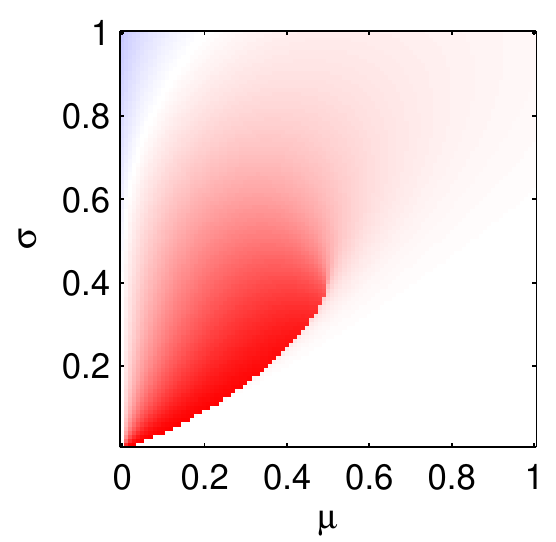}\\
\includegraphics[width=0.29\columnwidth]{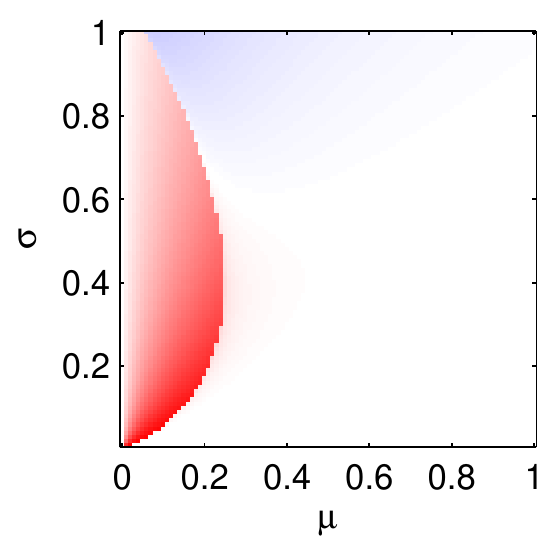}
\includegraphics[width=0.29\columnwidth]{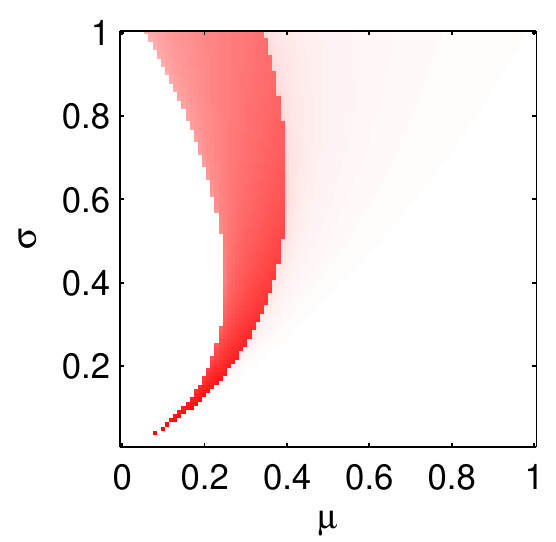}\\
\includegraphics[width=0.29\columnwidth]{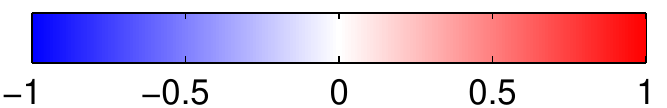} 
\caption{ Difference in fraction of failed nodes between class models.  Top Left:
  $X^{\ast}_\text{(i)}-X^{\ast}_\text{(ii)}$. Top Right:
  $X^{\ast}_\text{(i)}-X^{\ast}_\text{(iii)}$. Bottom Left:
  $X^{\ast}_\text{(ii)$\phi^0=0.25$}-X^{\ast}_\text{(iii)}$. Bottom Right:
  $X^{\ast}_\text{(ii)$\phi^0=0.4$}-X^{\ast}_\text{(ii)$\phi^0=0.25$}$.}
\label{fig:comparefailfrac}
\end{figure}

Figure \ref{fig:comparefailfrac} shows the difference in the final
fraction of failed nodes between the different model classes.  For
instance, the top left plot shows the difference between class (i) and
(ii), $X^\ast_\text{(i)} - X^\ast_\text{(ii)}$. It indicates that
constant load implies larger fraction of failures than load
redistribution, when the initial load is small ($\phi^0=0.25$). This,
however, does not hold for small $\mu$ and large $\sigma$, where more
nodes survive with constant
load. 
Overload redistribution leads to smaller systemic risk than constant load
(top right part of Figure \ref{fig:comparefailfrac}) and load
redistribution (bottom left part of Figure \ref{fig:comparefailfrac}),
except for very high $\mu$ and small $\sigma$. Interestingly, there is no
model class which leads to smaller systemic risk in the whole
$(\mu,\sigma)$-plane than the others.  Finally, the parameter $\phi^0$ in
the load redistribution model has a monotonic effect: The larger it is
the larger is the systemic risk (bottom right part of Figure
\ref{fig:comparefailfrac}).

Even under the basic assumption of a fully connected network, the
analysis carried out so far (denoted in the following as MF1) was able to
provide some insights in the relations among the three model classes. The
following remark is in order at this point. Mean field approaches have
known limitations. In principle, the foregoing analysis has little to say
about the outcome of individual realizations. For instance, consider the
top right corner of the top left plot of Fig. \ref{fig:failfrac}. In that
region, $X^\ast$ takes intermediate values around $0.25$. Does this imply
that one time every four there is a full breakdown and otherwise no
failure? Or, does it imply that in every realization one fourth of the
nodes fail? Notice that the cascade process is deterministic, and that
the sources of variability that are relevant for our purposes are the
initial distribution of net fragility $z(0)$ and the network structure
$A$. Our simulations show that, in absence of strong correlation in
$z(0)$ across nodes and in absence of strong modularity in the network,
the variability of $X^\ast$ across realizations is quite limited. A
robustness analysis is left as future work. For sure, in the regions in
which $X^\ast$ is very close to 1 the variability across realization is
negligible and this is a useful result in terms of systemic risk
estimation. Finally, a strong variability is expected, as usual, in the
proximity of the transition between small and large systemic risk.

The mean-field approach, could now be refined in various ways, in order
to take into account, for instance, the cases of non-fully connected
network, heterogeneous degree, or even degree-degree correlation,
following the methods that have been applied to epidemic spreading models
\cite{Pastor-Satorras.Vespignani2001EpidemicSpreadingin,Bogun'a2003Absenceofepidemic}. These
investigation are left as future work. 

In the remainder of this section, as an example, we analyse further the
constant load class on a network in which each node has on average $k$
neighbors (instead of $n-1$). This provides a first step to address the
influence of network topology on systemic risk. The fragility of a node
now takes values in the discrete set
$\{0,{1}/{k},{2}/{k},\dots,{j}/{k},\dots,1\}$. The probability of each
event corresponds to the probability that $j$ out of $k$ neighbours fail
at the same time, given that each fails with probability $q$. If failures
among neighbours of a node are independent, such probability follows a
binomial distribution $ B(j,k,q)$. We can further approximate the
probability $q$ that a node fails with the total fraction of failed nodes
$X$. Thus, we can write the probability density function for the
fragility as follows
\begin{eqnarray}
  p_{\phi(t)} &=& \sum_{j=0}^k B(j,k,X(t)) \, \delta_{\frac{j}{k}} \nonumber \\
  \text{with} \ B(j,k,X)&=&{j \choose k}
  X^j(1-X)^{k-j}.
\label{eq:phi}
\end{eqnarray}
It follows
\begin{align}
  p_{z(t)} &= \left(\sum_{j=0}^k B(j,k,X(t)
  \, \delta_{\frac{j}{k}}\right) \ast p_{-\theta}
\label{eq:FDrecusionK1} \nonumber \\
& = \sum_{j=0}^k B(j,k,X(t)) \, p_{-\theta+\frac{j}{k}} 
\end{align}
from which we can derive recursive equations in $X(t)$ analogous to
Eqns. (\ref{eq:recur:FD})--(\ref{eq:recur:OR}), to
compute $X^\ast$. A similar approach is used also in
\cite{Dodds2004Universalbehaviorin}. We denote this approach with MF2.

We can further refine the analysis by formulating a recursive equation
for the whole distribution $p_{z(t)}$ rather than for $X(t)$. This
approach can then take into account the fact that the distribution of
$z(t)$ is reshaped (and not simply shifted) during the dynamics.

For this purpose, we define the partial pdf of healthy nodes
$p^h_{z(t)} = \one_{[-\infty,0]}p_{z(t)}$, where $\one_{[a,b]}$
takes value one on the interval $[a,b]$ and zero elsewhere. The integral
of this function over the whole real axis gives the fraction of healthy
nodes, while the fraction of failed nodes is given by
\begin{align}
 X(t) = 1-\int_{-\infty}^0 p^h_{z(t)} dz.
\end{align}
Notice that the total mass of the function $p^h_{z(t)}$ is in
general smaller than one and decreases over time (therefore, strictly
speaking $p^h_{z}(t)$ is not a pdf). Because $p^h_{z}(t)$
only counts the healthy nodes, the \emph{fraction of currently failing}
(and not already failed!) nodes, $X_f$, is defined as
\begin{equation}
X_f(t) = \int_0^\infty p^h_{z(t)}(z)dz
\label{eq:failing}
\end{equation}
We can then write the recursive equation
\begin{align}
  p^h_{z(t+1)} &= \left(\sum_{j=0}^k
    B(j,k,X_f(t))\,\delta_{\frac{j}{k}}\right)
  \ast (\one_{[-\infty,0]} p^h_{z(t)})  \nonumber \\
  & = \sum_{i=0}^k B(j,k,X_f(t)) \one_{[-\infty,\frac{j}{k}]}
  p^h_{z(t)+\frac{j}{k}}.\label{eq:FDrecusionK2}
\end{align}
Summarizing, from Eqn. (\ref{eq:FDrecusionK2}) we can solve for the limit
distribution $p^{h\ast}_{z}$ or compute it numerically (after binning the
$z-$axis). This last method is denoted as MF3.

Methods MF2 and MF3 can be understood as conceptually different by
focussing on the net fragility of a single node coming for the
distribution.  For MF2 we compute the probability of a node to have a
certain net fragiity by its possibilities to have $0,1,2,\dots,k$ failed
neigbors, thus the maximum increase in fragility is $\tfrac{k}{k}$ over
all time steps. This fits to the inward agent-based dynamics, because we
focus on the receiving node which has $k$ in-neighbors. In MF3 instead,
after each time step the whole distribution of the net fragility is
reshaped. Thus, there is a nonzero probabiity that one node gets more
than $k$ increases in net fragility in successive time steps. We compute
how the fraction of currently failing nodes reshapes the distribution of
net fragility. Thus, we focus on the spreading node here and there is a
nonzero probability that one node can receive more than $k$ increases of
$\tfrac{1}{k}$ in two successive time steps, as it is also possible in a
network where in-degrees vary slightly. Thus, MF3 fits to the outwards
agent-based dynamics.

\begin{figure}
\centering 
\includegraphics[width=0.29\columnwidth]{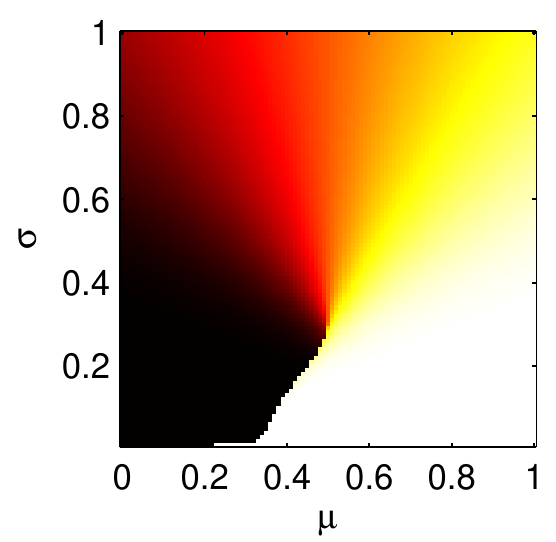}
\includegraphics[width=0.29\columnwidth]{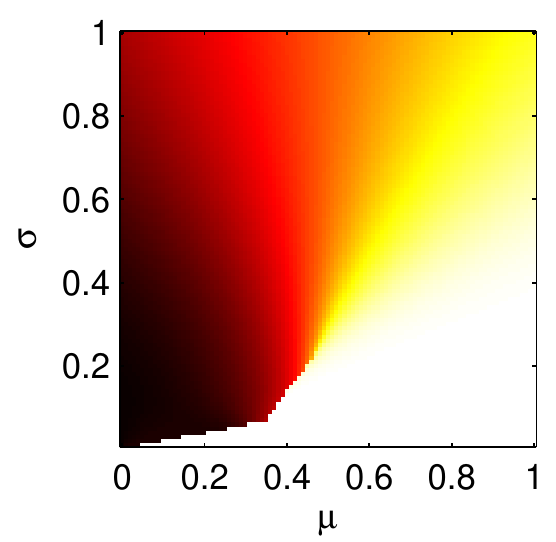}\\
\includegraphics[width=0.29\columnwidth]{nfigs/nfigMacroLegendHorizontal} 
\caption{Model class (i) 'constant damage', fraction of failed nodes 
$X^\ast$ on a regular graph with degree $k=3$. Plots are constructed 
in analogy to Figure \ref{fig:failfrac}. Left: mean field solution
from Eqn. \eqref{eq:FDrecusionK1} (MF2). Right: mean field
solution from Eqn. \eqref{eq:FDrecusionK2}(MF3). 
}
\label{fig:failfracFD12}
\end{figure}

\begin{figure}
\centering 
\includegraphics[width=0.29\columnwidth]{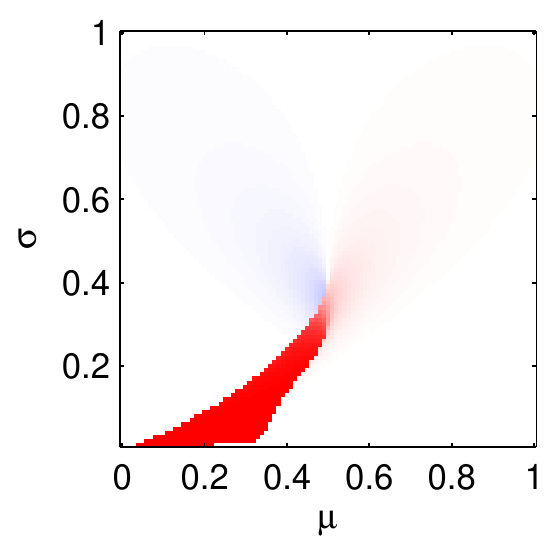}
\includegraphics[width=0.29\columnwidth]{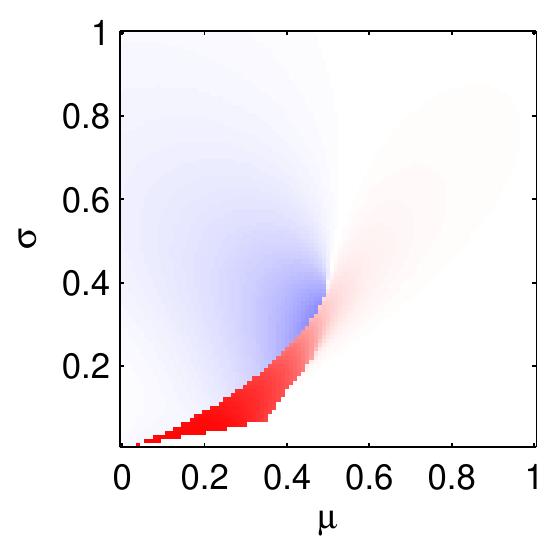}\\
\includegraphics[width=0.29\columnwidth]{nfigs/nfigMacroLegendBlueRedHorizontal} 
\caption{
  Model class (i) 'constant damage'. Difference between fraction of
  failed nodes $X^\ast$ on a regular graph with degree $k=3$ (shown in
  Figure \ref{fig:failfracFD12}) and on a fully connected network (shown
  in Figure \ref{fig:failfrac}), based on different mean field
  approaches. Left: plot of the difference
  $X^\ast_\text{MF2}-X^\ast_\text{MF1}$. Right: $X^\ast_\text{MF3}-X^\ast_\text{MF1}$.
  Color code as in Fig. \ref{fig:comparefailfrac}.
.}
\label{fig:difffailfracFD12}
\end{figure}

\begin{figure}
\centering 
\includegraphics[width=0.29\columnwidth]{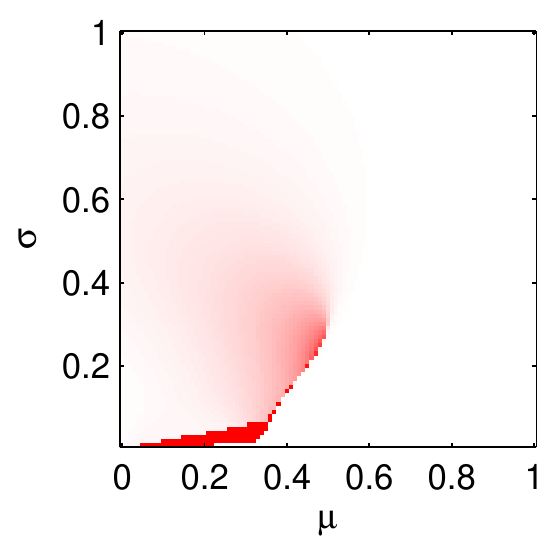}
\includegraphics[height=0.29\columnwidth]{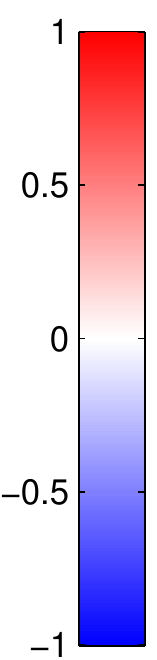} 
\caption{Model class (i) 'constant damage'. Difference
  $X^\ast_\text{MF2}-X^\ast_\text{MF3}$ between final fraction of failed nodes
  obtained with approaches MF2 and MF3 shown in Figure
  \ref{fig:failfracFD12}. Color code as in
  Fig. \ref{fig:comparefailfrac}.
  }
\label{fig:difffailfracFD12_2}
\end{figure}

Figures \ref{fig:failfracFD12}, \ref{fig:difffailfracFD12}, and
\ref{fig:difffailfracFD12_2} plot the limit fraction $X^\ast$ of failed
nodes in the $(\mu,\sigma)$ plane obtained from the recursive Equations
\eqref{eq:FDrecusionK1} (MF2) and \eqref{eq:FDrecusionK2} (MF3), as well
as a comparison with the case of fully connected network (MF1) and a
comparison between each other. Notice that, similar to MF1, we still
observe in both MF2 and MF3 a discontinuity line which vanishes as $\mu$
and $\sigma$ increase. The shape of the line varies in the three
analyses. In the third approach, MF3, the values of $X^\ast$ are
systematically smaller than in MF2.

Moreover, in MF1 the region of high systemic risk is less extended than
in the other approaches, although for intermediate values of
$\mu,\sigma$, values of $X^\ast$ in MF1 are larger than in MF2, MF3 (blue
regions in Fig. \ref{fig:difffailfracFD12_2}). This is due to the fact
that when many links are present, nodes are spreading the fragility more
evenly and so less failures take place, given the same initial
fragility. After the critical point the avalanche is larger.

On the other hand, approach MF2 always yields larger systemic risk than
MF3 which takes into account the whole distribution (see
Fig.~\ref{fig:difffailfracFD12_2}).  Thus, the inwards version of the
`constant load' model is more prone to systemic risk than the outward
version, this is especially relevant in the region of very low $\sigma$,
where MF2 shows full cascades up to $\mu\approx 0.3$, while MF3 is
already free from full cascades. A full cascade in MF3 is triggered only
for slightly higher $\sigma$.

\section{Generalization to Stochastic Cascading Models}
\label{sec:simple-cont-models}

\subsection{Stochastic description}\label{sec:stoch-descr}
In Section \ref{sec:form-model-cont} we have introduced a general model
of cascades based on a deterministic dynamics of the state $s_{i}(t)$ of
a node $i$, Eqn.(~\ref{eq:unify}), with a sharp transition from healthy
to failed state, at exactly $z_{i}=0$.
In this Section we propose a generalization of such process to a
stochastic setting. Interestingly, it will be possible to derive the
Voter Model as well as the stochastic contagion model SIS as particular
cases. This exercise will shed some new light on the connections between
cascade models and contagion models.

\begin{figure}[htbp] \centering
  \includegraphics[width=7cm,angle=0]{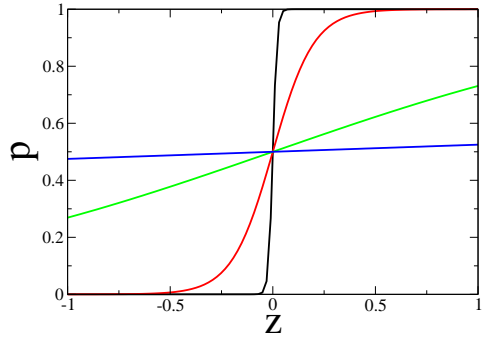}
  \caption{Probability $p=p(1)$, eqn. (\ref{eq:logit}), dependent on
    $z=z_{i}=z_{i}^{\prime}$ for several values of $\beta=\beta'$, to
    indicate the crossover from a random to a deterministic transition:
    (blue) $\beta=0.05$, (green) $\beta=0.5$, (red) $\beta=5$, (black)
    $\beta=50$ }
\label{fig:2}
\end{figure}

We assume that the failure of a node $i$ is a stochastic event occurring
with some probability dependent on the net fragility, $z_{i}(t)$, but
possibly also conditional to the current state, $s_{i}(t)$. We have in
mind a situation in which the probability to fail increases monotonically
as $z_i(t)=\phi_i-\theta_i$ becomes positive. Conversely, nodes can
switch from the failed state back to the healthy state and this is more
likely if $z'_i(t)=\phi_i-\theta'_i$ becomes negative. Notice that we
introduce an asymmetry, as $\theta'_i\not = \theta_i$ in
general. Compared to Equation (\ref{eq:unify}), now the dynamics is
defined as
\begin{eqnarray}
\label{eq:stoch-dynamics}
s_{i}(t+1)= \left\{
  \begin{array}{rcl}
    1 \; \,\mathrm{with}\, & p_{i}(1,t+1|1,t;z_i) & \mathrm{if} \ s_{i}(t)=1 \\
    1 \; \,\mathrm{with}\, & p_{i}(1,t+1|0,t;z_i) & \mathrm{if} \ s_{i}(t)=0 \\
    0 \; \,\mathrm{with}\, & p_{i}(0,t+1|0,t;z_i') & \mathrm{if} \ s_{i}(t)=0 \\
    0 \; \,\mathrm{with}\, & p_{i}(0,t+1|1,t;z_i') & \mathrm{if} \ s_{i}(t)=1 \\
\end{array}
\right.  
\end{eqnarray}
Here, $p_i(1,t+1|0,t;z_i)$ denotes the probability to find node $i$ in
state 1 at time $t+1$, conditional that it was in state 0 at time $t$,
etc. Obviously, 
\begin{eqnarray}
  \label{eq:hold}
1 &=& p_{i}(1,t+1|1,t;z_{i})+p_{i}(0,t+1|1,t;z'_{i}) \nonumber \\
1 &=& p_{i}(0,t+1|0,t;z'_{i})+p_{i}(1,t+1|0,t;z_{i})
\end{eqnarray}
In the following, we abbreviate the relevant conditional probabilities as
$p(1|0,z_{i})=p_{i}(1,t+1|0,t;z_{i})$, 
$p(0|1;z'_{i})=p_{i}(0,t+1|1,t;z'_{i})$ and denote them as transition
probabilities (per unit of time). Under Markov assumptions the
Chapman-Kolmogorov equation holds for the probability to find node
$i$ in state 1 at time $t+1$:
\begin{eqnarray}
  \label{eq:chap}
  p_{i}(1,t+1)&=&p_{i}(1,t+1|1,t,z_{i})\,p_{i}(1,t)
\nonumber \\ && +p_{i}(1,t+1|0,t,z_{i})\,p_{i}(0,t) 
\end{eqnarray}
With 
\begin{equation}
  \label{eq:conv}
 1 =  p_{i}(1,t)+p_{i}(0,t)
\end{equation}
and Eqn. (\ref{eq:hold}), this results in the dynamic equation
\begin{eqnarray}
  \label{eq:chap2}
    p_{i}(1,t+1)-\,p_{i}(1,t)&=&-p(0|1,z'_{i})\,p_{i}(1,t)
\nonumber \\ && +p(1|0,z_{i})\,\left[1-p_{i}(1,t) \right]
\end{eqnarray}
Stationarity, i.e. $p_{i}(1,t+1)-\,p_{i}(1,t)=0$, implies the so-called
detailed balance condition:
 \begin{equation}
  \label{eq:detailed}
   \frac{p_{i}(1)}{1-p_{i}(1)}=\frac{p(1|0;z'_{i})}{p(0|1;z_{i})}
\end{equation}
A very common assumption for $p_{i}(1)$ is the logit function: 
\begin{equation}
  \label{eq:logit}
  p_{i}(1;\beta,\beta';z_{i},z'_{i})=  \frac{\exp(\beta z_i)}{\displaystyle \exp(\beta
    z_i)+\exp(-\beta' z_i')}
\end{equation}
The parameters $\beta$, $\beta'$ measure the impact of stochastic
influences on the transition into the failed state and back into the
healthy state, accordingly. By varying $\beta$, $\beta'$ the
deterministic case ($\beta,\beta'\to \infty$) as well as the random case
($\beta,\beta'\to 0$) can be covered.  Figure \ref{fig:2} shows the
dependency of the probability $p=p(1)$ with respect to $\beta$ for the
symmetric case, $\beta=\beta'$, $z=z_{i}=z_{i}^{\prime}$.

The transition probabilities can be chosen in accordance with
Eqs. (\ref{eq:detailed}), (\ref{eq:logit}) as follows: 
\begin{align}
\label{eq:transition-prob}
p(1|0;z_{i})= &\gamma \frac{\exp(\beta z_i)}{\displaystyle \exp(\beta
  z_i)+\exp(-\beta' z_i')} \nonumber \\ 
p(0|1;z'_{i})= &\gamma' \frac{\exp(- \beta' z_i')}{\displaystyle
  \exp(\beta z_i)+\exp(-\beta' z_i')}
\end{align}
The parameters $\gamma$, $\gamma'$ set the range of the functions and
should be equal only if the detailed balance condition holds.  The
different thresholds $\theta$, $\theta'$ shift the position of the
transition from one state to the other.  The transition probabilities
thus depend on two sets of parameters, $\gamma$, $\beta$, $\theta$
characterizing the transition into the failed state, and $\gamma'$,
$\beta'$, $\theta'$ for the transition into the healthy state. These sets
differ in principle, but they play the same role in the transitions.

\subsection{Mean-field equations}
\label{sec:stoch-mean}

In analogy to Section \ref{sec:macr-reform}, we want to derive a dynamics
at the macro level for the expected fraction $X(t)$ of failed nodes at
time $t$. To this end, we start from the micro dynamics given by
Eqn. (\ref{eq:chap2}).
As a first mean-field assumption, we neglect correlations between
fragility and thresholds across nodes in the network. In other words,
we assume that the values of $z_i$ and $z_i'$ are drawn from the same
probability distribution $p_z(z(t))$, regardless of the identity of the
node. The expected change in the probability $p_i(1,z,t)$ for node $i$ is
obtained by integration:
\begin{eqnarray} 
  \label{eq:master2}
  E[p_i(1,t+1)&-&p_{i}(1,t)]=\int_\mathbb{R} p_z(z(t)) p(1|0;z) p_{i}(0,z,t) dz \nonumber\\
 & -& \int_\mathbb{R} p_z(z'(t)) p(0|1;z'_i) \,p_{i}(1,z',t)] dz'.
\end{eqnarray}
To avoid any confusion with the notation, we recall that $p_z$ is the
density function of the net fragility $z$, while $p(1|0;z)$ is the
probability that a node with net fragility $z$ switches from state $0$ to
state $1$, and finally $p_{i}(0,z,t)$ is the probability that node $i$ with net
fragility $z$ is in state $0$ at time $t$. 

We now average both sides of the equation above across nodes. In
particular, the average of the r.h.s. yields
\begin{align}
  \label{eq:master3}
  &\int_\mathbb{R} p_z(z(t)) p(1|0;z)\ \frac{1}{n}\sum_i p_{i}(0,z,t)dz  \nonumber \\
  & -\int_\mathbb{R} p_z(z'(t)) p(0|1;z')\ \frac{1}{n}\sum_i p_{i}(1,z',t)
  dz'.
\end{align}
Noticing that, for large $n$
\begin{eqnarray}
X(t) & = \frac{1}{n} \sum_i p_i(1,z,t)  \nonumber \\
1- X(t) & =\frac{1}{n}\sum_i p_{i}(0,z,t)
\label{eq:x}
\end{eqnarray}
we get
\begin{align}
  X(t+1)&-X(t) = (1-X(t)) \, \int_\mathbb{R} p_z(z(t))
  p(1|0;z(t)) dz \nonumber \\
&  - X(t) \, \int_\mathbb{R} p_z(z'(t)) p(0|1;z') dz'.
  \label{eq:master4}
\end{align}
Equation (\ref{eq:master4}) describes the dynamics of the expected
fraction of failures in a system with both heterogeneity of threshold,
$\theta$, or fragility, $\phi$, and with stochasticity in the cascading
mechanism. We can now obtain mean-field equations for various existing
models, by specifying (1) the transition probabilities $p(1|0;z)$ and
$p(0|1;z')$, and (2) the distribution $p_{\theta}(\theta)$ for the
thresholds $\theta_i$.

\subsection{Recovering Deterministic Cascade Models}
\label{recover}
In order to recover the deterministic models of Section
\ref{sec:specific-cascade-models}, we first notice that in those cases
the transition from state $s=0$ to $s=1$ is not really conditional to the
state at previous time. Actually, in these models a node changes to a
certain state with a probability which is independent of its current
state. We emphasize, however, that our framework is general enough to
cover cases in which failure is really conditional on $s$.

For the models discussed in Section \ref{sec:specific-cascade-models}, we
can assume $\theta'=\theta$ and thus $z'=z$, and further $\beta=\beta'$,
$\gamma=\gamma'=1$. We have then
\begin{eqnarray}
  \label{eq:indep}
  p(1|0;z) &=& p(1|1;z) =  p(1;z) 
  \nonumber \\ 
  p(0|1;z) &=& p(0|0;z) = p(0;z).
\end{eqnarray}
We now set $\beta\to \infty$, which implies that the transition
probability in Equation (\ref{eq:transition-prob}) tends to the Heaviside
function:
\begin{equation}
p(1|0;z) = \Theta(z)\;;\quad p(0|1;z)= \Theta(-z)
\label{eq:heavy}
\end{equation}
Since, for any real function $g$ holds
\begin{eqnarray}
\int_\mathbb{R} g(x) \Theta(x) dx = \int_0^{\infty} g(x) dx,  
\end{eqnarray}
we obtain
\begin{align}
  \label{eq:master7}
 X(t+1)- & X(t)=  (1-X(t))\, \int_0^{\infty} p_z(z(t)) dz \nonumber \\
& - X(t)\,\int_{-\infty}^0 p_z(z(t)) dz.
\end{align}
Because of $\int_{-\infty}^0 p_z(z(t)) dz + \int_0^{\infty}p_z(z(t)) dz =
1$, this finally yields
\begin{align}
  \label{eq:master8}
& X(t+1)= \int_0^{\infty}p_z(z(t)) dz
\end{align}
Eqn. (\ref{eq:master8}) coincides with Eqn. (\ref{eq:Xmacro}).

\subsection{Recovering Stochastic Models with Homogeneous Threshold}
\label{recover-stoschastic}
In order to recover models of herding and stochastic contagion, we
instead keep the stochastic nature of the failure but we assume that the
failure threshold is the same across nodes, $\theta_i=\theta$, $\forall
i$. In a mean field approximation, we replace the individual fragility
with the average one, so that also $z_i$ is constant across the nodes
$z_i=z \, \forall i$. Then the probability density of $z$ in Equation
(\ref{eq:master4}) is equivalent to a delta function and the integral
over $dz$ drops. The macroscopic mean-field equation then reads
\begin{align}
  \label{eq:master5}
  X(t+1)-X(t)=(1-X(t)) \, p(1|0;z) - X(t) \, p(0|1;z)
\end{align}
Eqs. (\ref{eq:master5}) will be the starting point for discussing
specific contagion models in Sections \ref{sec:voter}, \ref{sec:sis}.

\subsection{Voter Model}
\label{sec:voter}

The linear voter model (LVM) is a very simple model of herding behavior.
The dynamics is given by the following update rule: a voter, i.e. a node
$i\in A$ of the network, is selected at random and adopts the state of a
randomly chosen nearest neighbor ${j}$. After $n$ such update events,
time is increased by 1. The probability to choose a node $j$ in state 1
from the neighborhood of node $i$ is proportional to the frequency of
nodes with state 1 in that neighborhood, $f_{i}$ (and conversely for
state 0). Consequently, the transition probability towards the opposite
state is proportional to the local frequency of the opposite state. It is
also independent of the current state of the node.
\begin{align}
  \label{eq:VM-fi}
  p(1)=p(1|1)&=p(1|0)=f_{i} \nonumber \\
  p(0)=p(0|0)&=p(0|1)= 1-f_i
\end{align}
In order to match this dynamics within our framework, we consider values
of $\beta$ of the order of 1. From Equation (\ref{eq:transition-prob}),
 we obtain in linear approximation:
\begin{eqnarray}
  \label{eq:logit-lin}
   p(1|0,z_{i}) &=& \frac{\gamma}{2}\left[1+\beta z_i\right] \nonumber \\
   p(0|1,z_{i}) &=& \frac{\gamma'}{2}\left[1-\beta' z'_i\right].
\end{eqnarray}
With $\theta_{i}=\theta_{i}'$, $\beta=\beta'$, $\gamma=\gamma'$, this
matches the transition probabilities for the LVM provided that:
\begin{eqnarray}
  \gamma \left[1+\beta (\phi-\theta_{i})\right] &=&2f_{i} \nonumber \\
  \gamma \left[1-\beta (\phi-\theta_{i})\right] &=&2\left(1-f_{i}\right) 
\end{eqnarray}
This is realized by choosing
\begin{equation}
\gamma=1\;;\  \beta=2\;;\ \theta=\frac{1}{2} \;\Rightarrow \ \phi_i=f_i
\end{equation}
We note that the threshold $\theta$ coincide with the unstable
equilibrium point of the LVM, $1/2$, that distinguishes between
minorities and majorities in the neighborhood.  The fragility equals the
local frequencies $f_{i}$ of infected nodes, and does not depend on the
node itself. If a majority of nodes in the neighborhood has failed, this
more likely leads to a failed state of node $i$; if the failed nodes are
the minority, this can lead to a transition into the healthy state.
Since the fragility coincides with fraction of neighbours in state 1 (or
0), VM fits in the first model class described in
Sec. \ref{sec:models-with-constant-damage}, with the specificity that the
failure process is stochastic and the threshold is homogeneous across
nodes. 

As a consistency check, if we assume for one moment the failure
process to be deterministic, one could directly apply
Eqn. (\ref{eq:recur:FD}). Since the probability distribution of the
threshold would be trivially a delta function $\delta_{1/2}(\theta)$, its
cumulative distribution would be the Heaviside function
$P_{\theta}(X)=\Theta(X-1/2)$. This would imply that the dynamics reaches
$X^*=1$ as stable fix point as soon as $X(0)>1/2$ and viceversa for
$X^*=0$.

Coming back to the usual stochastic VM, in order to obtain the mean-field
dynamics, we now approximate $f_i(t)$ with $X(t)$, i.e. we replace
$p(1|0,z) = X(t)$, $p(0|1,z) = 1- X(t)$ in Equation
(\ref{eq:master5}). This recovers the well known mean-field dynamics of
the LVM, ${dX}/{dt}=0$, i.e. the expected asymptotic fraction of failures
(which differs from the individual realizations) coincide with the
initial fraction $X(0)$ of failed nodes.

With a similar procedure we can also account for nonlinear VM
\cite{Schweitzer2009Nonlinearvotermodels:}, in which the probability to
switch to a failed or healthy state is a non-linear function of the
fraction of failed nodes in the neighborhood:
\begin{align}
  \label{eq:VM_trans_prob1}
  p(1|0;z_{i})=p(1)= & f_{i}(t)\; F_1(f_{i}(t)) \nonumber \\
  p(0|1;z_{i})=p(0)= & (1-f_{i}(t))\; F_2(f_{i}(t))
\end{align}
$F_1$ and $F_2$ are frequency dependent functions which describe the
non-linear response of a node on the fraction of failed nodes in the
neighborhood. If we again replace $f_{i}$ with the global frequency of
failures $X$, we arrive at the macroscopic dynamics in mean-field limit:
\begin{align}
  \label{eq:masterVM}
  X(t+1)-X(t) = (1-X(t))\,X(t)\;\left[F_1(X)-F_2(X)\right]
\end{align}
In the linear case $F_1=F_2=1$, the prediction for the expected value of
$X$ does not give sufficient information about individual realizations of
the Voter dynamics. In fact, it is well known that the global outcome of
the LVM leads to global failure with a probability equal to the initial
fraction of infected nodes, $X(0)$. In other words, if we run a
simulation with e.g. $X(0)=0.3$ for 100 times, then in 30
cases we will reach a state of global failure, whereas in 70
cases, no failure at all will prevail. This differs from the case of the
cascading models described in Sect. \ref{sec:specific-cascade-models}, in
which the mean field dynamics gives us some more information about
individual realizations.

In the non-linear case ($F_1\not =1$, $F_2\not =1$), different scenarios
arise depending on the nonlinearity. In
\cite{Schweitzer2009Nonlinearvotermodels:} it was shown that even a small
non-linearity may lead to either states where global failure is always
reached, or to states with a coexistence of failed and healthy nodes. It
is worth noticing that both of these scenarios are obtained with
\emph{positive frequency dependence}, i.e. a transition probability to
fail that increases monotonically with the local frequency $f$. Thus,
small deviations in the nonlinear response can either enhance systemic
risk, or completely prevent it.

\subsection{SIS-SI model}
\label{sec:sis} 

The SIS model
\cite{Vespignani.Pastor-Satorras2002EpidemicSpreadingScale-free} is the
most known model of epidemic spreading. On the microlevel, healthy nodes
get infected with probability $\nu$ if they are connected to one or more
infected nodes. In other words, the parameter $\nu$ measures the
infectiousness of the disease in case of contact with an infected
node. This means that the effective transition probability of node $i$
from healthy to infected state is proportional to the probability $q$
that a neighbour is infected times the degree $k_i$ of the node. Indeed,
the larger the number of contacts, the more likely it is to be in contact
with an infected node. On the other hand, failed nodes recover
spontaneously with probability $\delta$. The transition probabilities are
then as follows:
\begin{equation}
  \label{eq:tans-si}
  p(1|0,z_{i})= \nu \, k_i \,q; \;\; p(0|1) = \delta
\end{equation} 
with $0\leq \nu \leq 1$. We do not redefine, as usually done, the
infection rate as $\lambda=\nu/\delta$ with $\delta=1$, because we want to
cover the case $\delta=0$, as we will see below. 

We interpret of course infection state as failure state. Matching the
transition probabilities of SIS with the ones in our framework, we
obtain:
\begin{eqnarray}
  \gamma \left[1+\beta (\phi-\theta_{i})\right] &=&2  \nu \, k_{i} \, q \nonumber \\
  \gamma' \left[1-\beta' (\phi-\theta'_{i})\right] &=& 2 \delta
\end{eqnarray}
This implies that  our framework recovers the transition probabilities of
the SIS model, provided that: 
\begin{eqnarray}
\gamma=1 &\;;\ & \beta=2\;;\ \theta=\frac{1}{2} \;\Rightarrow \
\phi_i=\nu \, k_{i} \, q
\nonumber \\
\gamma'=2 \delta&\;;\ &  \beta'=0
\end{eqnarray}

In order to understand the relation of SIS with the other models, we can
approximate the probability $q$ that a node fails with the fraction of
failed neighbours $f_i$. The resulting expression for the fragility,
$\phi_i=\nu k_{i} f_i$, is proportional to the fraction of failed nodes
as in model class (i) of
Sec. \ref{sec:models-with-constant-damage}. However, the term $k_i$
implies that the infection probability grows with the number of
connections in the network. This feature makes the biggest difference
between the SIS model and the cascade models studied in the previous
sections, apart of course from the fact that the contagion process is
stochastic and the threshold homogeneous. Another important feature that
emerges is the asymmetry in the transition probabilities between healthy
and failed state and backwards.

In order to derive a macroscopic dynamics, we apply the mean-field
approximation $f_{i}\sim q \sim X$ and we assume a homogeneous network
with $k_i=k$ for all nodes. Starting from Equation (\ref{eq:master5}), we
obtain
\begin{eqnarray}
  \label{eq:SIS-growth}
X(t+1)-X(t) & = \nu \, k\, X(t) (1-X(t)) - \delta X(t) 
\end{eqnarray}
The last negative term in the R.H.S. of Eqn. (\ref{eq:SIS-growth})
implies that there is no global spreading of infection if $\nu < \nu_c=
\frac{\delta}{k}$ and the only stable fix point is $X^\ast=0$. For $\nu
\ge \nu_c$ there is a unique stable fix point with $X^\ast>0$.

As it is well known, the existence of a critical infection rate $\nu_c$
does not hold, however, if, instead of the mean-field limit with
homogeneous degree, a heterogeneous degree distribution of the nodes is
assumed
\cite{Vespignani.Pastor-Satorras2002EpidemicSpreadingScale-free}. The
implications of degree heterogeneity and degree-degree correlation in
epidemic spreading have been investigated in a number of works
\cite{Bogun'a2003Absenceofepidemic}.

The SI model, in which no transition into the healthy state is possible,
is recovered setting additionally $\delta=0$. We then obtain the logistic
growth equation:
\begin{align}
  \label{eq:si-growth}
X(t+1)-X(t) = \nu \,k \;X(t) (1-X(t))
\end{align}
where $X(t)=1$ is the only stable fix point of the dynamics.  Any initial
disturbance of a healthy state eventually leads to complete infection.

We conclude by noting that, despite its simplicity, the SI model has been
used to describe a number of real contagious processes, such as the
spread of innovations \cite{Jackson.Rogers2007RelatingNetworkStructure}
or herding behavior in donating money \cite{schweitzer2008epidemics}. In
the latter case, the mean-field interaction was provided by the mass
media. In other words, because of the constant and homogeneous information
about other people's donations, the individual transition depends on the
global (averaged) frequency of donations instead of the local
one. Interestingly, it could be shown that in the particular example of
'epidemic' donations, the time scale depends itself on time, indicating a
slowing down of the dynamics due to a decrease in public interest.

\section{Conclusion}

In this paper, we wish to clarify the meaning and the emergence of
systemic risk in networks with respect to several existing models. To
unify their description, we propose a framework in which nodes are
characterized (1) by a discrete failure state $s_{i}(t)$ (healthy or
failed) and (2) by a continuous variable, the \emph{net fragility},
$z_{i}(t)$, capturing the difference between fragility $\phi_{i}$ of the
node and its failing threshold $\theta_{i}$.  By choosing an appropriate
definition of fragility in terms of the failure state and/or the
fragility of neighbouring nodes, we are able to recover, as special cases,
several cascade models as well as contagion models previously studied in
the literature.

Our paper contributes to the investigation of these models in several
ways. First of all, we have provided a novel framework to cover both
cascade and contagion models in a deterministic approach, which is
further suitable to be generalized also to the stochastic case. Secondly,
our framework allows us to unify a number of existing, but seemingly
unrelated models, pointing out to their commonalities and
differences. Thirdly, we are able to identify three different model
classes, which are each characterized by a specific mechanism of
transferring fragility between different nodes.  These are (i) 'constant
load', (ii) 'load redistribution', and (iii) 'overload redistribution'.

Systemic risk, within our framework, is defined as the stable fraction of
failed nodes $X^\ast$ in the system.  As $X^\ast= 1$ denotes the complete
breakdown of a system, we are interested in trajectories of the system
where $X^\ast$ is much below one. In order to determine these
trajectories, in this paper we derive a \emph{macroscopic dynamics} for
$X(t)$ based on the microscopic dynamics.  As a major contribution of
this paper, we are able to find, for each of the three classes, macroscopic
equations for the final fraction of failed nodes in the mean-field limit.

In order to compare the systemic risk between the three classes, we have
studied the macroscopic dynamics of each of them with the same initial
conditions. Most importantly, we found that the differences on the
microscopic level translate into important differences on the macroscopic
level, which are visualized in a phase diagram of systemic risk. This
indicates, for each of the model classes, which given initial conditions
result into what total fraction of failure. This way we could verify
that, for instance, in class (ii) there is a first-order transition
between regions of high systemic risk and regions with low systemic
risk. In contrast, class (i) displays such a sharp transition only in
some smaller part of the phase space, while class (iii) does not display
any abrupt transition at all. Such an insight helps us to understand
whether and for which parameters small variations of initial conditions
may lead to an abrupt collapse of the whole system, in contrast to a
gradual increase of systemic risk.

In addition to the macro dynamics, we have also investigated the
different model classes on the microscopic level. A number of network
examples made clear how the different transfer mechanisms affect the
microdynamics of cascades. As an interesting insight, we could
demonstrate that the role of nodes with high degree change depending on
type of load transfer. In the inwards variant of first class model, high
degree nodes are more protected from contagion, whereas in the outwards
variant of the same class, they become more exposed to contagion if they
are connected to many low degree nodes (which holds for disassortative
networks). Furthermore, we could point out that the results strongly
depend on the position of the initially failing node. A systematic
analysis identifying the crucial nodes from a systemic risk point of view
is left for future work.

Finally we have extended our general framework so to encompass models of
stochastic contagion, as known from VM and SIS. Both of these models
belong to the first class, but differently from the models studied in
Section \ref{sec:models-with-constant-damage}, the threshold is
homogeneous and the failure is stochastic. Hence, it becomes more clear
how these established models of herding behavior and epidemic spreading
are linked to the 'cascade' models discussed in the literature.

Our work could be extended in several ways. First, one could apply
techniques to deal with heterogeneous degree distributions to the three
classes of cascading model, introduced in
Sec. \ref{sec:specific-cascade-models}. This could be carried out also in
the presence of degree-degree correlation, as recently discussed for
contagion models \cite{Bogun'a2003Absenceofepidemic}.
Furthermore, one could investigate more in detail the case of both
heterogeneous threshold and stochastic failures. Compared to the simple
SIS, it seems more realistic to assume that the probability of contagion
depends on an intrinsic heterogeneous property of the nodes (the
threshold). Such heterogeneity could also play a crucial role, as it has
been found for the heterogeneity in the degree
\cite{Vespignani.Pastor-Satorras2002EpidemicSpreadingScale-free}.

A last remark is devoted to the discussion of systemic risk. In our
paper, we have provided mean-field equations to calculate the total
fraction of failed node in a system, which we regard as a measure of
systemic risk. This implies that systemic risk is associated with a
system state of global failure, i.e. there is no 'risk' anymore, as
almost all nodes already failed. In contrast, it could be also
appropriate to define 'systemic risk' as a situation, where the system
has not failed yet, but small changes in the initial conditions or
fluctuations during the evolution may lead to its complete collapse. Our
general framework has already contributed insight into this problem, by
identifying those areas in the (mean-field) phase diagram where we can
observe a sharp transition between a globally healthy and a globally
failed system. This is related to precursors of a crisis as it identifies
parameter constellations to make a system vulnerable that looks
apparently healthy. On the other hand, using our approach we were able to
assess that for certain transfer mechanisms such an abrupt change in the
global state is not observed at all -- which means that systems operating
under some conditions are less vulnerable to small changes.

\subsection*{Acknowledgements}
This work is part of a project within the COST Action P10 ``Physics of
Risk''. J.L. and F.S. acknowledge financial support from the Swiss State
Secretariat for Education and Research SER under the contract number
C05.0148. S.B. and F.S. acknowledge financial support from the ETH
Competence Center 'Coping with Crises in Complex Socio-Economic Systems'
(CCSS) through ETH Research Grant CH1-01-08-2.


\begin{appendix}

\section{Eisenberg-Noe model}\label{sec:eisenbergnoe}

An interesting model of contagion which has not been investigated
in the econophysics literature is the one developed by Eisenberg and Noe
\cite{Eisenberg.Noe2001SystemicRiskin}. It introduces a so called
fictitious default algorithm as a clearing mechanism in a financial system
of liabilities. When some agents in the system cannot meet fully their
obligations, the task of computing how much each one owes to the other
becomes nontrivial in presence of cycles in the network of
liabilities.

The basic assumptions of the clearing mechanism are (i) limited
liability (a firm need not spend more than it has), (ii) absolute
priority of debt over cash (a firm has to spend all available cash to
satisfy debt claims first), (iii) no seniority (all claims have the
same priority).

A financial system of $n$ firms is described by a vector of total
obligations $x^0$, a matrix of relative liabilities $A$, and a vector of
operating cash flows $\theta$.  $x^0_i$ is the total amount of
liabilities firm $i$ has towards other firms, $a_{ij}$ specifies what
fraction of its own total obligations firm $i$ owes to firm $j$, and
$\theta_i$ determines the liquid amount of money of firm $i$. Thus,
$a_{ij}x^0_i$ is the nominal liability $i$ has to $j$. The matrix
$A$ is row-stochastic, which means all entries are non-negative and rows
sum up to one. This condition ensures that individual obligations sum up
to the total obligations $x_j^0$. The expected payments to firm $i$ from
its debtors is thus:
\begin{equation}
  \label{eq:oblig}
(A^Tx^0)_i=\sum_{j=1}^na_{ji}x^0_j
\end{equation}
If it happens that the total cash flow, i.e., the expected repayments of
others plus operating cash-flow, is less than the total obligations, i.e.
\begin{equation}
\theta_i + \sum_{j=1}^na_{ji}x^0_j < x^0_i 
\label{eq:cash}
\end{equation}
firm $i$ cannot meet its obligation in full and defaults. This implies a
reduction of the expected payments to its creditors, which might in turn
default as a second-order effect, and so on. This makes this model close
to the class of overload redistribution because the expected payments of
a node do not vanish entirely when it fails.

The fictitious default algorithm defined in
\cite{Eisenberg.Noe2001SystemicRiskin} consists of finding a clearing vector
$x^\ast$ of total payments which fulfills the equation
\begin{equation}
 x_i = \min\{  \theta_{i} + \sum_{j=1}^n a_{ji} x_j \;,\;  x_i^0 \}
\end{equation}
for all $i$. As shown in \cite{Eisenberg.Noe2001SystemicRiskin}, using
mild assumptions, such a clearing vector exists and is unique and the
fictitious default algorithm, with $x(0) = x^0$, is well defined. The
sequence $x(t)$ represents a decreasing sequence of clearing vector
candidates which terminates in at most $n$ steps at the clearing
vector. The new clearing vector candidate $x(t+1)$ is computed from a
given candidate $x(t)$ taking into account the first order defaults given
clearing vector candidate $x(t)$, but not the second order
defaults. These are checked in the successive time steps.

Following our general framework presented in Section
\ref{sec:form-model-cont}, we can define fragility as
\begin{equation}
 \phi_i(x(t),x^0,A) = x_i^0 - \sum_{j=1}^na_{ji}x_j(t) \label{eq:ENfragility}
\end{equation}
which is the amount of debt which has to be covered by the operating cash
flow $\theta_i$, given the current candidate for the clearing vector
$x(t)$. From $x(t)$, $x^0$, $A$, and $\theta$ we can determine the
failing state $s_{i}(t+1)$ as in Equation (\ref{eq:unify}) as
\begin{equation}
 s_i(t+1) = \Theta(\phi_i(x(t),x^0,A) - \theta_i).
\end{equation}
Given a clearing vector candidate $x(t)$, the \emph{value of the equity}
of firm $i$ is given by
\begin{equation}
 \theta + \sum_{j=1}^n a_{ji} x_j(t) - x_i(t) 
\end{equation}
which is the operating cash flow plus the expected amount of payments
received by others minus the payment to others which are possible, given
the currently expected payments from the other.  A new clearing vector
candidate $x(t+1)$ is computed from $x(t)$ by determining the failing
state $s(t+1)$. This leads to a simple fix point equation (see
\citet[p. 243]{Eisenberg.Noe2001SystemicRiskin}), which usually has a
unique fix point. That means, the fictitious default algorithm is
constituted in such a way that it solves a system of linear equations.
If successful, the algorithm runs until a clearing vector is found which
gives that the value of the equity is zero (and not negative) for a firm
in default, and positive for a non-defaulting firm. At least one
non-defaulting firm should be found by the fictitious default algorithm,
which means that there is at least one firm $i$ for which it holds
$x_i(t) = x^0_i$ holds. If not, then the clearing vector candidates
diverge toward $-\infty$, and the algorithm fails. This represents a full
break down of the financial system.

The relation between this model and our third model class is not
straightforward because the new clearing vector candidate $x(t)$ is not
necessarily always uniquely defined by the current fragility $\phi(t)$ as
given in \eqref{eq:ENfragility}. Therefore, we chose to study the simpler
models in Sec. \ref{sec:models-with-redistr-overload}.

Investigating the macro-perspective as in Section \ref{sec:macr-reform},
one finds that the Eisenberg-Noe model can be approximated by the
macroscopic equation for the overload redistribution. The approximations
would be fairly good when the system is close to a fully connected
network (everybody borrows equally from everybody else) and uncorrelated
operating cash flows.

\section{Further examples}
\label{sec:further-examples}
Section \ref{sec:specific-cascade-models} has pointed out that the
propagation of cascades, in addition to the mechanism of transfer,
strongly depends on the initial condition, in particular on the position
of the first failing node. In order to further illustrate this important
point, we present additional examples with a different initially failing
node. All these examples start from the setup shown in Figure
\ref{fig:All:init}. Their outcome should be compared to the respective
examples discussed in Figures \ref{fig:FDfail3}, \ref{fig:FBfail3},
\ref{fig:FBOfail3}
\begin{figure}[h]
\centering
\includegraphics[width=0.45\columnwidth]{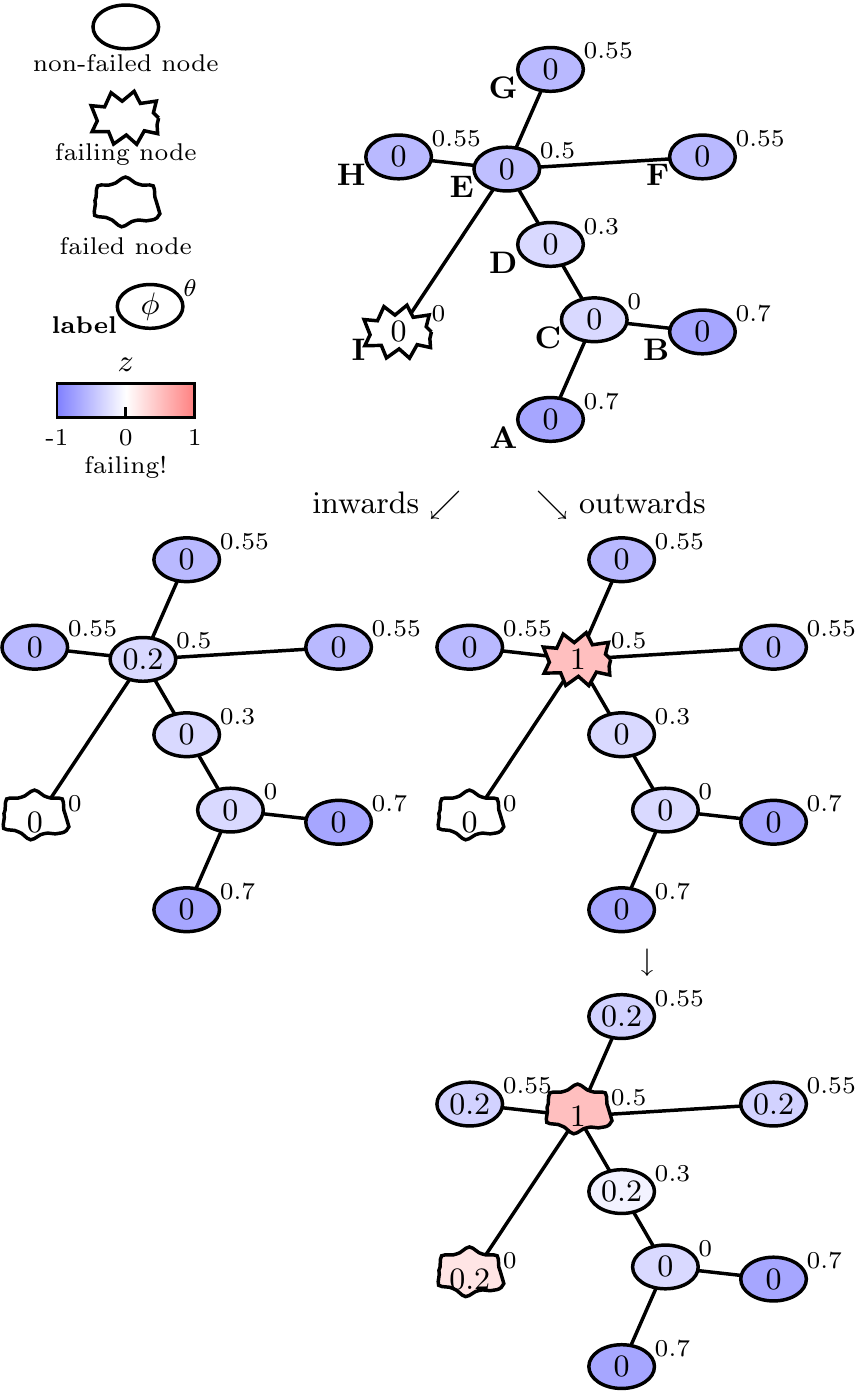}
\caption{{\bf Constant damage.} Example to be compare with Figure
  \ref{fig:FDfail3}. Here, node {\bf I} initially
  failes.}\label{fig:FDfail9}
\end{figure}

\begin{figure}
\centering
\includegraphics[width=0.42\columnwidth]{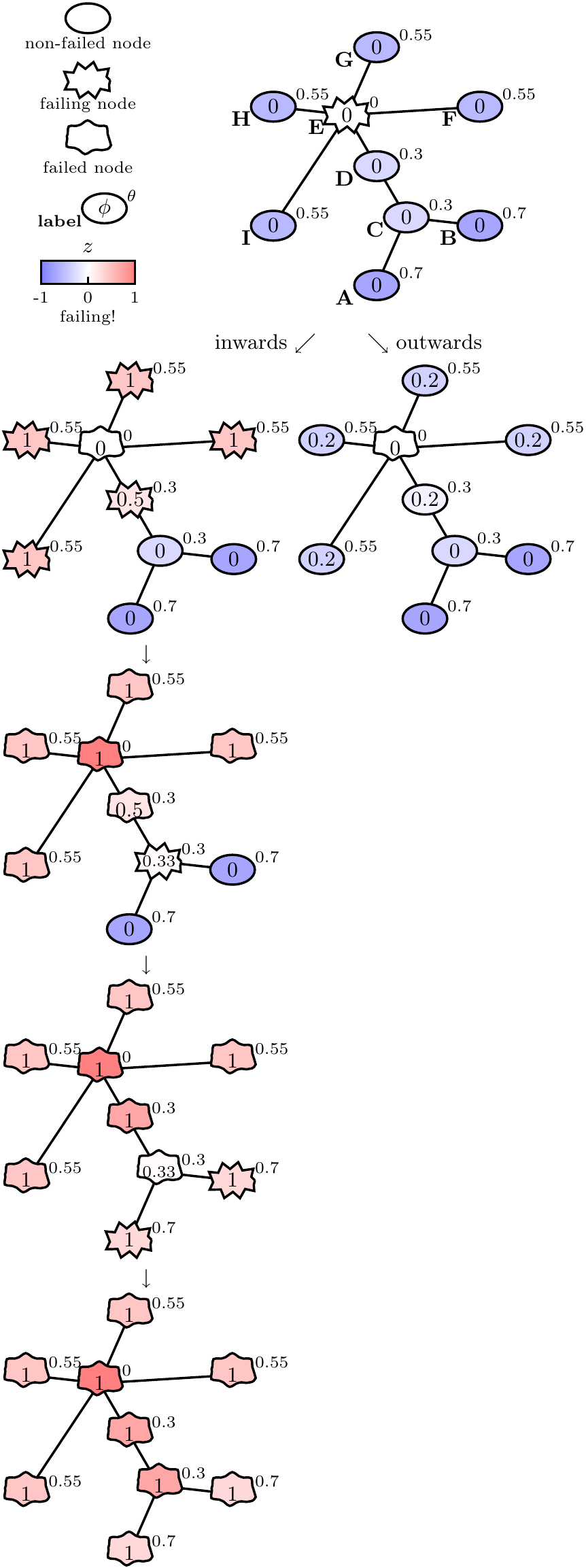}
\caption{{\bf Constant damage.} Example similar to the one in Figure
  \ref{fig:FDfail9} but with highest degree node {\bf E} failing
  initially. The example clearly shows a difference in the spreading
  properties of hubs: in the `inwards variant' the hub spreads failures
  to low-degree nodes; the opposite for the `outwards
  variant'.}\label{fig:FDfail5}
\end{figure}

\begin{figure}[htbp]
\centering
\includegraphics[width=0.45\columnwidth]{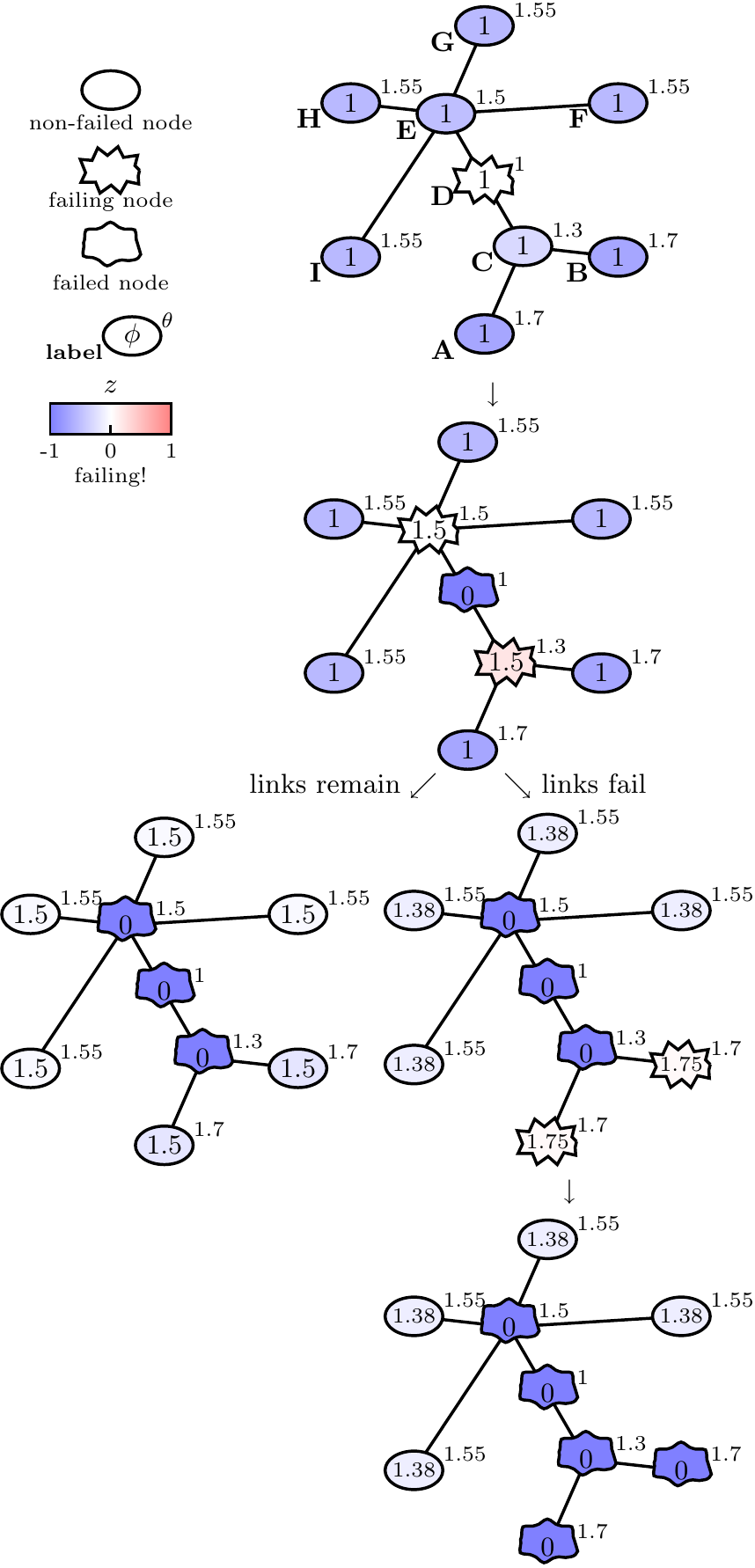}
\caption{{\bf Load redistribution.} Compare with
  Fig. \ref{fig:FBfail3}. Here, node {\bf D} fails initially. w.r.t to
  the previous case, this leads to a propagation of failures in the
  opposite direction, in the LLSC variant (links fail), while nothing
  changes in the LLSS variant (links remain).}\label{fig:FBfail4}
\end{figure}

\begin{figure*}[htbp]
\centering
\includegraphics[width=0.9\textwidth]{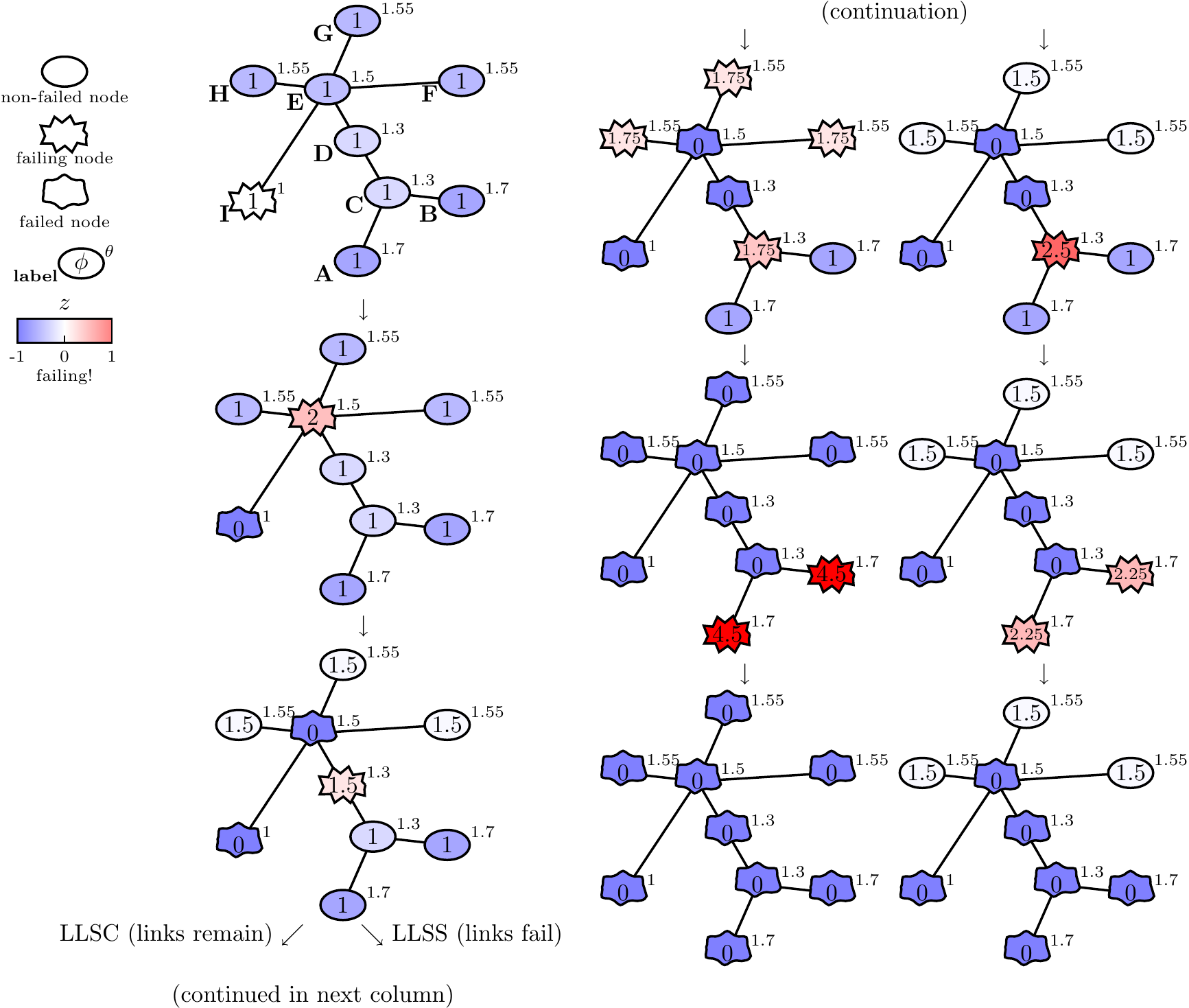}
\caption{{\bf Load redistribution.} Compare with Figure
  \ref{fig:FBfail3}. Here, node {\bf I} fails initially. This leads to
  full breakdown in the LLSC variant (links remain), while some nodes do
  not fail in the LLSS variant (links fail) because of a disconnection in
  the network.}\label{fig:FBfail9}
\end{figure*}

\end{appendix}
 
\end{document}